\documentclass[apj]{emulateapj}

\usepackage{amsmath}
\usepackage[usenames,dvipsnames]{color}
\usepackage{amssymb}

\newcommand{\epsbold}{\mbox{\boldmath$\epsilon$}}
\newcommand{\thetabold}{\mbox{\boldmath$\theta$}}

\newcommand{\B}{\textsc{Bayes-ME}}
\newcommand{\BL}{\textsc{Bayes-LTE}}

\newcommand{\Bpar}{B_\parallel}
\newcommand{\Bperp}{B_\perp}

\newcommand{\didlnoi}{\frac{\partial I(\lambda)}{\partial \lambda}}

\newcommand{\didltwonoi}{\frac{\partial^2 I(\lambda)}{\partial \lambda^2}}
\newcommand{\coschi}{\cos 2 \chi}
\newcommand{\sinchi}{\sin 2 \chi}

\begin{document}

\title{Model selection for spectro-polarimetric inversions}
\author{A. Asensio Ramos\altaffilmark{1,2}, R. Manso Sainz\altaffilmark{1,2}, M. J. Mart\'{\i}nez Gonz\'alez\altaffilmark{1,2}, 
B. Viticchi{\'e}\altaffilmark{3,4}, D. Orozco Su\'arez\altaffilmark{5}, H. Socas-Navarro\altaffilmark{1,2}}
\altaffiltext{1}{Instituto de Astrof\'{\i}sica de Canarias, 38205, La Laguna, Tenerife, Spain}
\altaffiltext{2}{Departamento de Astrof\'{\i}sica, Universidad de La Laguna, E-38205 La Laguna, Tenerife, Spain}
\altaffiltext{3}{ESA/ESTEC RSSD, Keplerlaan 1, 2200 AG Noordwijk, The Netherlands}
\altaffiltext{4}{Dipartimento di Fisica, Universit\`a degli Studi di Roma ``Tor Vergata'', via della Ricerca Scientifica 1, 00133 Rome, Italy}
\altaffiltext{5}{National Astronomical Observatory of Japan, Mitaka, Tokyo 181-8588, Japan}
\email{aasensio@iac.es}

% % abstract cannot exceed 300 words
\begin{abstract}
Inferring magnetic and thermodynamic information from spectropolarimetric observations
relies on the assumption of a parameterized model atmosphere
whose parameters are tuned by comparison with observations. 
Often, the choice of the underlying atmospheric model is based on subjective reasons.
In other cases, complex models are chosen based on objective reasons (for instance, the
necessity to explain asymmetries in the Stokes profiles) but it is not clear what degree of complexity
is needed. The lack of an objective way of comparing models has, sometimes, led to opposing
views of the solar magnetism because the inferred physical scenarios
are essentially different. We present the first
quantitative model comparison based on the computation of the Bayesian evidence ratios for spectropolarimetric observations.
Our results show that there is not a single model appropriate for all profiles simultaneously. Data
with moderate signal-to-noise ratios favor models without gradients along the line-of-sight.
If the observations shows clear circular and linear polarization signals above the noise level, models with
gradients along the line are preferred. As a general rule, observations with large signal-to-noise
ratios favor more complex models. We demonstrate that the evidence ratios correlate well
with simple proxies. Therefore, we propose to calculate these proxies when
carrying out standard least-squares inversions to allow for model comparison in the future.
\end{abstract}

\maketitle

%%%%%%%%%%%%%%%%%%%%%%%%%%%%%%%%%%%%%%%%
%%%%%%%%%%%%%%%%%%%%%%%%%%%%%%%%%%%%%%%%
% INTRODUCTION
%%%%%%%%%%%%%%%%%%%%%%%%%%%%%%%%%%%%%%%%
%%%%%%%%%%%%%%%%%%%%%%%%%%%%%%%%%%%%%%%%
\section{Introduction}
Spectropolarimetry is a very powerful diagnostic technique. It has allowed us to study in depth 
the thermodynamical and magnetic
properties of the solar and stellar plasmas. However, the valuable information encoded in the 
Stokes profiles is often difficult to extract. The response of the spectral shape of a given spectral
line to changes of the properties of the plasma is very convoluted, non-linear
and, in many occasions, non-local.

In spite of the complex relation between the physical parameters and the
emergent Stokes profiles, several simple diagnostic tools have been developed in the past 
and are still under wide use in solar physics. Among them, the line ratio technique \citep{stenflo73,stenflo10,stenflo11},
the center-of-gravity method \citep{semel70,rees_semel79} and the application of calibration curves 
between spectral line and magnetic field properties \citep[e.g.,][for recent applications]{lites08,imax11} have had special relevance.

During the last few decades we have witnessed the
development and systematic application of nonlinear inversion codes. They extract physically
relevant information by comparing the observed Stokes profiles to those
synthesized in appropriate atmospheric models. Because of the non-linearity between the physical
parameters and the observables, these inversion methods make use of elaborate 
time-consuming non-linear optimization methods. The first heroic (because of the low
computational power available at that time) efforts made use of relatively simple
physical models, of which the Milne-Eddington (ME) approximation is the
most widely spread \citep[e.g.,][]{harvey72,auer_heasly_house77,landi_landolfi04}. Although the
simplifying assumptions made by the ME approximation may not be fully fulfilled in real
solar plasmas, it is still one of the most widely used models, in part, because of its
analytical simplicity. Even state-of-the-art inversion codes such as VFISV \citep{borrero07,borrero_vfisv10},
used for inferring magnetic field vectors from the Helioseismic and Magnetic Imager (HMI; onboard
the Solar Dynamics Observatory) data, or as MILOS \citep{orozco_hinode07} and
MERLIN \citep{skumanich_lites87,lites07}, currently applied to data from the Hinode spacecraft,
are based on this assumption.

The increase in computational power made it feasible to use more elaborate
models. A fundamental leap forward was the application of the idea of response
functions \citep{landi_response77} to the inversion of Stokes profiles with non-trivial
depth stratifications of the physical quantities. The first representative
of this family of codes was SIR \citep[Stokes Inversion based on Response functions;][]{sir92}.
The presence of gradients along the line-of-sight (LOS) of the physical
properties are of importance for explaining the strong asymmetries observed in magnetized regions 
\citep{solanki88,grossman_doerth88,solanki93,sigwarth99,khomenko03,marian08,viticchie_2_11}.
Based on the same strategy, \cite{socas_navarro98} developed a code capable of dealing
with lines in NLTE (non-local thermodynamical equilibrium). This model
has been mainly applied for the inversion of Ca \textsc{ii} infrared triplet
lines, which are formed under strong NLTE conditions \citep[e.g.,][]{socas_trujillo_ruiz00,jaime10}.

Models based on the assumption of microstructured magnetic atmospheres which incorporate small-scale
fluctuations of the physical quantities along the LOS have also been proposed
\citep{sanchez_almeida_misma97}. A property of such models is the natural
generation of asymmetric profiles, which has been proposed as an
alternative to explain the observed asymmetries \citep{sanchez_almeida89,sanchez_almeida00,viticchie_1_11}.

The history of inversion codes displays an interesting characteristic. The complexity
of the models proposed has increased with time thanks to two factors.
First, the computational power has allowed us to solve the non-linear optimization
problem faster. Second, the quality of spectropolarimetric observations has improved
with time, with instruments systematically achieving signal-to-noise ratios above
1000 and reaching 10$^4$ in many cases. This forces the inversion code 
to fit minute variations of the shape of the Stokes profile. In many cases, such
small modifications of the shape of the profiles are encoding important physical effects.

Although counterintuitive, the availability of high-quality observations has also intensified
a discussion that has been present from the beginning of the history of inversion codes.
The selection of the model to be used for the inversions is often influenced by 
subjective reasons. A model can be chosen deliberately to be simple because of
the necessity of inverting large maps as fast as possible. This is the case of
HMI \citep{borrero_vfisv10}. An inversion code based on a given model 
is sometimes chosen based on the availability of the inversion code and the
expertise to use it. A model might be chosen based on its capability to
fit as much detail of the Stokes profiles as possible. Sometimes the
selection of a model is driven by spatial resolution. This is the case
when there is a clear indication in the spectrum that structures are not 
resolved. Other possibilities can be invoked, but they sometimes contain 
a large subjective content.

The choice of the model is also important because it critically affects the interpretation of the
inferred parameters. This is specially relevant for parameters that are more indirectly related to
observables, like the magnetic field of non-resolved structures. There are widespread examples 
in the literature, especially when signals are weak. For instance, there has been some controversy 
regarding the magnetic field strength in the quiet Sun inferred from infrared or visible
Fe \textsc{i} lines \citep[e.g.,][]{khomenko03,lites_socas04,dominguez06,marian08}.
More recently, conflicting properties of the
magnetic field in internetwork regions of the quiet Sun have been obtained by \cite{lites08}
and \cite{orozco_hinode07} on the one hand, and by \cite{stenflo10} on the other,
from the very same Hinode observations \citep[and partially supported with data
from ground by][]{beck_rezaei09}. These discrepancies are only ``apparent'' because 
they are caused by the application of different modelings when the
organization of internetwork magnetic fields is still an unknown. 
Once we know the nature of the magnetic fields in the quiet Sun or, in 
other words, we have access to the most probable model among all possible ones,
the results will be reliable. 
This problem does not arise for parameters that are more directly related to observables 
like bulk velocities, field azimuths, magnetic flux, etc.

This paper does not deal with the actual implementation of particular inversion
algorithms and codes, or their respective efficiencies, but with the suitability
of the underlying physical models to explain the observations.
To this aim we perform the first fully Bayesian comparison of models
for the inversion of Stokes profiles. To this end, we analyze a sample of Stokes profiles
observed with the spectropolarimeter \citep[SP;][]{lites_hinode01} 
aboard Hinode \citep{kosugi_hinode07} and with the vector magnetograph
IMaX \citep{imax11} onboard the Sunrise balloon \citep{sunrise10}.
% and with
% the Tenerife Infrared Polarimeter \citep[TIP;][]{collados_tipII07}. 
We discuss, for
different cases, the model (chosen from a pool of fixed models) that is favored by data and
their relative probabilities based on the computation of the evidence. The evidence
is calculated using the Bayesian inference code developed by
\cite{asensio_martinez_rubino07} and recent extensions that we have made to
the code to deal with models with gradient.

\section{Model selection theory}
\label{sec:model_comparison}
We apply model selection theory to determine which is the model best suited 
for explaining the Stokes profiles observed in a pixel \citep[e.g.,][for a general
description and more details]{trotta08}.
Let's assume we have $N_\mathrm{mod}$ models $\{ \mathcal{M}_i,i=1\ldots N_\mathrm{mod}\}$ competing 
to explain the same set of observations formally represented by $D$. 
Here, $D$ will be represented by a formal vector $\mathbf{d}=[I(\lambda_1), I(\lambda_2), ..., Q(\lambda_1), ..., U(\lambda_1), ..., V(
\lambda_1), ...]$ whose elements are the values of the Stokes parameters $I$, $Q$, $U$, and/or $V$ at certain wavelengths $\lambda_1, 
\lambda_2, ...$.
By a \emph{model} we mean an algorithm that depends on a set of $N_j^{(i)}$ parameters $\thetabold_i=(\theta_{i; 1}, \theta_{i; 2}, 
..., \theta_{i; N_{j}^{(i)}})$ (often, the temperature at one or several points in a model atmosphere; the magnetic field strength, in
clination, and azimuth; the density, etc), whose output is a prediction $\mathbf{y}(\thetabold_i)$ of the data. 
The Bayes theorem \citep{jaynes03,mackay03,gregory05} states that the posterior probability of each model at the light of the obs
erved data is
\begin{equation}
p(\mathcal{M}_i|D) = \frac{p(D|\mathcal{M}_i) p(\mathcal{M}_i)}{p(D)},
\label{eq:prob_model}
\end{equation}
where $p(\mathcal{M}_i)$ is our prior belief in each model (which we will assume to be the 
same for all the models considered here; see below), while $p(D)$ is just a normalization constant:
\begin{equation}
p(D) = \sum_{i=1}^{N_\mathrm{mod}} p(D|\mathcal{M}_i) p(\mathcal{M}_i).
\end{equation}
Finally, $p(D|{\cal M}_i)$ is the evidence or marginal likelihood, which is the key 
ingredient of our model comparison, and is given by the following integral \citep[e.g.,][]{trotta08,asensio_spw6_11}:
\begin{equation}
p(D|\mathcal{M}_i) = \int \mathrm{d}\thetabold_i p(\thetabold_i|\mathcal{M}_i) p(D|\thetabold_i,\mathcal{M}_i).
\label{eq:evidence}
\end{equation}
The quantity $p(\thetabold_i|\mathcal{M}_i)$ is the prior distribution for the model parameters.
If, for example, we assume that $\theta_{i;j}$ (i.e., the $j$-th parameter of mode $i$) is \emph{a priori} distributed
uniformly between $\theta_{i;j}^\mathrm{min}$ and $\theta_{i;j}^\mathrm{max}$, then
\begin{equation}
p(\theta_i^j|\mathcal{M}_i) = \left\{
\begin{array} {clc}
\left[ \theta_{i;j}^\mathrm{max} - \theta_{i;j}^\mathrm{min} \right]^{-1} & & \theta_{i;j}^\mathrm{min} < \theta_{i;j} < \theta_{i;j}^\mathrm{max} \\
0 & & \mathrm{otherwise}
\end{array}
\label{eq:prior_uniform}
\right..
\end{equation}
The quantity $p(D|\thetabold_i,\mathcal{M}_i)$ in Eq. (\ref{eq:evidence}) is the likelihood, which is computed from the observed data. 
Assuming that the observations are corrupted with uncorrelated Gaussian random noise, then
\begin{equation}
p(D|\thetabold_i,\mathcal{M}_i) = \prod_{j=1}^M \left(2\pi \sigma_j^2\right)^{-1/2} \exp \left[ -\frac{ \left(y_j(\thetabold_i)-d_j \right)^2}{2\sigma_j^2} \right],
\end{equation}
\citep[for mode details, see][]{asensio_martinez_rubino07,asensio_hinode09}.
In general, we will assume our priors to be uniform (Eq. \ref{eq:prior_uniform}), hence, the evidence 
in Eq. (\ref{eq:evidence}) simply reads:
\begin{equation}
p(D|\mathcal{M}_i) = \left[ \prod_{j=1}^{N_i} \frac{1}{(\theta_i^j)^\mathrm{max} - (\theta_i^j)^\mathrm{min}} \right] 
\int_{\Omega} \mathrm{d}\thetabold_i p(D|\thetabold_i,\mathcal{M}_i),
\label{eq:evidence_simplified}
\end{equation}
where the integral is computed extended over the volume $\Omega$ which contains the region where the priors are
non-zero.
It is important to note that, if a model parameter is completely unconstrained by the observed data
(so the ensuing likelihood does not depend on this parameter), the evidence does not penalize it
because it factorizes from the integral.

Given two models, $\mathcal{M}_0$ and $\mathcal{M}_1$ that are proposed to explain
an observation, the ratio of posteriors
\begin{equation}
\frac{p(\mathcal{M}_0|D)}{p(\mathcal{M}_1|D)} = \frac{p(\mathcal{M}_0)}{p(\mathcal{M}_1)} \frac{p(D|\mathcal{M}_0)}{p(D|\mathcal{M}_1)},
\end{equation}
is used to compute how more probable one model is with respect to the other \citep{jeffreys61}. The
ratio of evidences is known as the Bayes factor, $B_{01}$, and it is trivially given by:
\begin{equation}
B_{01} = \frac{p(D|\mathcal{M}_0)}{p(D|\mathcal{M}_1)}.
\end{equation}
If both models are assumed to have the same a-priori probability (which is what we have
assumed in all subsequent computations), the ratio of posteriors
is just the Bayes factor. Large values of $B_{01}$ indicate a preference for model $\mathcal{M}_0$
while small values indicate a preference for model $\mathcal{M}_1$. Table \ref{tab:jeffreys_scale}
gives the modified Jeffreys scale that can be used to translate values of the Bayes
factor into strengths of belief \citep{jeffreys61,kass_raftery95,gordon_trotta07}.

%%%%%%%%%%%%%%%%%%%%%%%%%%%%%
\begin{table}
\caption{Modified empirical Jeffreys' scale \citep[taken from][]{trotta08}\protect\footnotemark}
\label{tab:jeffreys_scale}
\centering
\begin{tabular}{c c c}
\hline\hline
$|\ln B_{01}|$ & Odds & Strength of evidence\\
\hline
$<1.0$ & $\lesssim 3:1$ & Inconclusive\\
$1.0$ & $\sim 3:1$ & Weak evidence\\
$2.5$ & $\sim 12:1$ & Moderate evidence\\
$5.0$ & $\sim 150:1$ & Strong evidence\\
\hline
\end{tabular}
\end{table}
%%%%%%%%%%%%%%%%

Therefore, model comparison is a matter of computing $p(\mathcal{M}_i|D)$ for all
models and calculating their ratios for pairs of models. As shown by Eq. (\ref{eq:evidence}),
the evidence contains a balance between the quality of the fit (encoded in the
likelihood) and the number of parameters (encoded on the prior).
In order to gain some insight consider the following example. Let ${\cal M}_1$ be 
a model with a single parameter $\theta$, and ${\cal M}_0$ another one with this 
parameter fixed to the value $\theta_0$. If the prior on the parameter for 
${\cal M}_1$ is flat with a range $\Delta\theta$ sufficiently large to acommodate the likelihood, then
\begin{eqnarray}
p(\theta|\mathcal{M}_0) &=& \delta(\theta-\theta_0) \nonumber \\
p(\theta|\mathcal{M}_1) &=& 
\left\{
\begin{array} {clc}
\frac{1}{\Delta \theta} & & \theta_{i;j}^\mathrm{min} < \theta_{i;j} < \theta_{i;j}^\mathrm{max} \\
0 & & \mathrm{otherwise}
\end{array}
\right.
\end{eqnarray}
Now, let the likelihood be a relatively peaked function around the 
value $\bar{\theta}$, with a characteristic width $\delta\theta$, then, 
from Eq. (\ref{eq:evidence}) the evidences in both cases read
\begin{eqnarray}
p(D|\mathcal{M}_0) &=& p(D|\theta_0,\mathcal{M}_1) \nonumber \\
p(D|\mathcal{M}_1) &\approx& p(D|\hat{\theta},\mathcal{M}_1) \frac{\delta \theta}{\Delta \theta},
\end{eqnarray}
Assuming identical priors for both models, the evidence ratio is
\begin{equation}
\frac{p(D|\mathcal{M}_1)}{p(D|\mathcal{M}_0)} \approx \frac{p(D|\hat{\theta},\mathcal{M}_1)}{p(D|\theta_0,\mathcal{M}_0)} \frac{\delta \theta}{\Delta \theta}.
\end{equation}
Consequently, since the ratio of likelihoods has to be larger than or equal to 1 (model $\mathcal{M}_1$
contains model $\mathcal{M}_0$), model $\mathcal{M}_1$ is preferred to model $\mathcal{M}_0$ when the prior space is 
not so large with respect to the width of the likelihood. This shows that the evidence ratio 
works as a Bayesian \emph{Occam's razor}. 

The main difficulty in model comparison
is that the evidence in Eq. (\ref{eq:evidence}) is computationally very demanding because it is the result of a high-dimensional
integral \citep[e.g.,][and references therein]{trotta08}. In recent years, some efficient algorithms, 
specially those based on nested sampling \citep{skilling04}, have been developed to deal with this problem. The codes
that we use in this paper make use of the Multinest algorithm 
\citep{feroz09} which performs very well in our cases.

\section{Observations}
\subsection{Spectropolarimetry}
We select the profiles for the application of the Bayesian model comparison for
spectropolarimetric data from a representative sample of what one can find in the
solar photosphere. For the quiet Sun, they have been extracted from the 
observations analyzed by \cite{lites08}. They were obtained at disk center 
on 2007 March 10 with the spectropolarimeter SOT/SP aboard Hinode with a
spatial resolution of $\sim0.32''$.
The observed spectral region consists of the Fe \textsc{i} doublet at 6301.5 and 6302.5 \AA\ with
a spectral resolution close to 3$\times$10$^5$, resulting in a total of 112 
wavelength points. After calibration, the standard deviation of the
noise in Stokes $Q$, $U$ and $V$ in units of the continuum is estimated to be of the order of 1.1-1.2$\times$10$^{-3}$.
Likewise, the standard deviation of Stokes $I$ computed on a continuum window
is estimated to be $\sim$6$\times$10$^{-3}$, the increase being probably produced by 
flat-fielding effects.
Given the large computational effort that the estimation of the evidence requires, we
have focused on individual profiles extracted from the Stokes $V$ classification calculated by \cite{viticchie_1_11}
using a k-means unsupervised classification algorithm widely used in machine learning \citep{everitt95,bishop06}.
We have selected individual profiles from the observations whose 
polarization amplitude is, in any of the Stokes parameters,
above a threshold of 4.5 times the standard deviation of the noise. This way, we avoid large uncertainties
in the model parameters as pointed out by \cite{asensio_hinode09}. The considered profiles
are shown in black curves in Fig. \ref{fig:map_fits} and their classification, including the
nomenclature for the shape of the profiles used by \cite{viticchie_1_11} is displayed in 
Table \ref{tab:evidences}. Of the six groups available (network, blue-lobe, red-lobe, asymmetric, 
antisymmetric and $Q$-like), we have picked representative profiles, some of them 
having no apparent linear polarization above the noise threshold and some of them
showing clear signals. We consider that they represent a good sample of what
one can find in the quiet Sun observed with Hinode.

Concerning the profiles associated to umbra and penumbra, they were extracted
from Hinode observations obtained on 2007 February 27. The estimated noise
level in the umbra for Stokes $Q$, $U$ and $V$ is of the order of 5$\times$10$^{-3}$ in
units of the continuum intensity, while it increases to 0.02 for Stokes $I$. The noise level is much larger than
for the quiet Sun profiles, partly because of the reduced number of photons and also
because of the apperance of molecular lines that we do not fit.

\subsection{Imaging polarimetry}
Four IMaX observations have been chosen to
study model comparison in imaging polarimetric data. The first one is characterized
by Stokes $V$ above the noise level and Stokes $Q$ and $U$ below the noise. 
The second one has Stokes $Q$ and $U$ above the noise and $V$ below. The third
has all Stokes parameters above the noise, while the fourth has all Stokes parameters
below the noise. The analysis of
model comparison for the inversion of these profiles is of special 
relevance given their low spectral sampling. The observations
consist of the four Stokes parameters at $-80$, $-40$, $+40$, $+80$ and $+227$ m\AA\ around
the Fe \textsc{i} line at 5250.209 \AA. The spectral point spread function (PSF)
has very extended tails, typical of Fabry-P\'erot instruments. Instead of taking
it into account exactly, we follow \cite{lagg_imax10} who substituted the real PSF by 
a Gaussian PSF of 85 m\AA\ ($\sim 2.9$ km s$^{-1}$) of full width 
at half-maximum (FWHM). Although the Gaussian PSF does not present extended
tails, its convolution with the FTS spectrum gives line profiles similar
to those obtained using the correct PSF \citep[see][]{imax11}.

The estimated noise level is 4$\times$10$^{-3}$ for
Stokes $I$ and 10$^{-3}$ for Stokes $Q$, $U$ and $V$, all in units of the 
continuum intensity. The spatial resolution of the instrument
is estimated to be between 0.15$''$ and 0.18$''$ \citep{lagg_imax10}.

Table \ref{tab:properties_profiles} shows some interesting properties of the
observed Stokes profiles: $P_\mathrm{tot}$ is the maximum total polarization,
$P_V$ is the maximum circular polarization, $a$ and $A$ are the area and
amplitude asymmetries respectively \citep[e.g.,][]{solanki_stenflo86}, while $v_\mathrm{V}$ and $v_\mathrm{I}$
are the Stokes $V$ zero-crossing velocity and the Stokes $I$ velocity of the
minimum. The velocities are related to the rest wavelength of the spectral
lines (they are not absolute). The sign of the 
area asymmetry is chosen equal to the
sign of the bluest peak of Stokes $V$, following \cite{martinezpillet97}. We have not tabulated
the asymmetries for the IMaX data because we are not confident on their values with
only 5 points in wavelength and of class 34 because it clearly shows the presence
of several lobes making the definition of asymmetries invalid.

\section{Atmospheric models}
Among the infinitely many models one might build to reproduce the emergent
Stokes profiles from a magnetized atmosphere, we consider for our analysis those
most widely used in the literature. 
All of them are based on different approximations for the solution of
the radiative transfer equation for polarized radiation in a plane-parallel atmosphere \citep[see][]{landi_landolfi04}:
\begin{equation}
\frac{d \mathbf{S}}{dz} = \epsbold - \mathbf{K} \mathbf{S},
\end{equation}
where $z$ is the spatial coordinate along the ray, $\mathbf{S}=(I,Q,U,V)^t$ is the Stokes vector, $\epsbold$ the emissivity
vector, and $\mathbf{K}$ the propagation matrix.

All models considered assume local thermodynamic equilibrium. Therefore,
$\epsbold$ and $\mathbf{K}$ depend on the local thermodynamic and magnetic
properties of the medium but not on $\mathbf{S}$ itself. 
Three types of hypotheses are assumed for the variation of $\epsbold$ and
$\mathbf{K}$ with $z$: i) constant along the line of sight, ii) some
thermodynamic quantities vary (linearly) but the magnetic field is constant,
iii) all quantities may vary with $z$. 
Finally, we consider models with a single atmosphere occupying the whole
element, and models with two independent atmospheres within the same resolution
element. 
In the latter case, the emergent Stokes profiles are given by
\begin{equation}
\mathbf{S} = f \mathbf{S}_1 + (1-f) \mathbf{S}_2,
\label{eq:filling_factor}
\end{equation}
where $\mathbf{S}_1$ and $\mathbf{S}_2$ are the emergent profiles of each of the
two components, which occupy fractions $f$ and $(1-f)$ of the pixel,
respectively.
Two possibilities are considered for these two components. First, $\mathbf{S}_1$
corresponds to a magnetic component (hence, $Q, U$, and $V$ are, in general,
non-zero), while $\mathbf{S}_2$ forms in a field-free atmosphere (hence,
$Q=U=V=0$).
Second, both atmospheres are magnetized.

In these models, we consider the formation of the widely used Fe~{\sc i} 630~nm
doublet and the 5250.2 \AA\ line.
The atomic data for the synthesis of the lines has been compiled from the VALD database
\citep{vald_piskunov95,vald_kupka99}. The data is summarized in Table \ref{tab:atomic}. 
Collisional broadening is treated under 
the formalism of \cite{anstee_omara95} for the broadening 
of spectral lines (only for allowed transitions), with the velocity parameter ($\alpha$) and the 
line broadening cross section ($\sigma$, in units of $a_0^2$, with $a_0$ the Bohr radius) obtained from the code
developed by \cite{barklem98}.

Polarimetric data for the imaging polarimetry case is calculated from high spectral resolution synthetic
profiles convolved with Gaussian transmission functions of 85~m\AA\  width, then
sampled at the five spectral positions used in the V5-6 mode of IMaX. 

In order to limit the scope of our analysis, 
we left out of our study physical models considering the formation of the
spectral line in non-LTE conditions \citep[e.g.,][]{socas_trujillo_ruiz00}. It is well established that
these effects play a negligible role in the formation of the spectral features
we are interested in \citep{shchukina_trujillo01}.
We did not consider, either, models with more than two components \citep[e.g.,][]{bernasconi96,beck07,beck_rezaei09} or
statistical models in which the thermodynamic and/or magnetic properties of the
atmosphere are given statistically \citep[e.g.,][]{sanchez_almeida_misma96,sanchez_almeida_misma97,carroll07}. 
These models have been introduced to explain some of the most complex observed features which are difficult or
impossible to explain with the models considered here. Furthermore, we also limit
this analysis to the Zeeman effect, neglecting atomic level polarization. The study of these
important class of models are left for a further study.

\subsection{Weak-field approximation}
The first model we use for this analysis is the well-known weak-field approximation \citep[][]{landi73}.
It is valid whenever the splitting produced in a given spectral line via the Zeeman effect 
is smaller than the intrinsic line broadening. In this approximation, there is a very
simple relation between the magnetic properties of the plasma and the derivatives of the Stokes
$I$ profile and the emergent Stokes parameters at
first order for Stokes $V$ and second order for Stokes $Q$, $U$ (see App. A).
It is important to point out that, strictly speaking, the likelihood in the weak-field approximation
depends only on $D=\{Q,U,V\}$. Stokes $I$ enters only through the computation of the
derivatives used in Eq. (\ref{eq:circ_lin_pol}) and, consequently, is part of the
model, not of the observations\footnote{When noisy quantities are part of the model, the
likelihood function has to be modified accordingly \citep{gregory05}. 
\cite{asensio_manso11} presented such an example for the inversion of
Stokes profiles with a model with local stray-light.}. Because of this, the formalism of model selection cannot be directly applied to compare
the weak-field model with the rest of models, because they have to share the same 
set of observations $D=\{I,Q,U,V\}$. To fix this issue, 
we propose a slightly revised weak-field approximation in which we also model Stokes $I$
as an absorption line at central wavelength $\lambda_0$ using:
\begin{equation}
I(\lambda) = 1 - d H\left( \frac{\lambda-\lambda_0-\lambda_0 v_\mathrm{LOS}/c}{\Delta \lambda_\mathrm{dopp}}, a \right),
\label{eq:stokesi_wf}
\end{equation}
where $H(v,a)$ is a Voigt profile. Therefore, each Stokes $I$ is defined with the aid of the 
line absorption ($d$), the Doppler width of the line in wavelength units 
($\Delta \lambda_\mathrm{dopp}$), the wavelength shift due to a macroscopic 
bulk velocity ($v_\mathrm{LOS}$) and the damping constant ($a$).
To these parameters, we add the LOS component of the magnetic field vector ($\Bpar$), the
projection of the magnetic field vector on the perpendicular to the LOS ($\Bperp$) and 
the azimuth of the magnetic field ($\chi$). 
This approximation to the weak-field approximation is also interesting for the V5-6 mode of IMaX 
because of the scarcity of points. This way, the wavelength derivatives of Stokes $I$ needed
for defining circular and linear polarization profiles are evaluated with more precision.

%%%%%%%%%%%%%%%%%%%%%%%%%%%%%
\begin{table*}[t]
\caption{Atomic parameters.}
\label{tab:atomic}
\centering
\begin{tabular}{c c c c c c c c}
\hline\hline
$\lambda$ (\AA)  & Excitation & $\log(gf)$ & Transition & $\sigma$ & $\alpha$ & $\bar{g}$ & $\bar{G}$ \\
& pot. (eV) & & & & & &\\
\hline
6301.498 & 3.654 & $-0.745$ & $^5$D$_2$ - $^5$P$_2$ & 839.9 & 0.243 & 1.667 & 2.517 \\
6302.494 & 3.687 & $-1.203$ & $^5$D$_0$ - $^5$P$_1$ & 856.9 & 0.240 & 2.500 & 6.250 \\
5250.209 & 0.121 & $-4.938$ & $^5$D$_0$ - $^7$P$_1$ &  -  & - & 3.000 & 9.000 \\
% 15648.509 & 5.426 & $-0.675$ & $^7$D$_1$ - $^7$D$_1$ & 977.0 & 0.229 & 3.000 & 9.000 \\
% 15652.874 & 6.246 & $-0.043$ & $^7$D$_5$ - $^7$D$_4$ & 1445.0 & 0.330 & 1.500 & 2.247 \\
\hline
\end{tabular}
\end{table*}
%%%%%%%%%%%%%%%%

%%%%%%%%%%%%%%%%%%%%%%%%%%%%%
% \begin{deluxetable*}{ccccccccc}
% \tabletypesize{\scriptsize}
% \tablecaption{Model parameters and prior ranges.}
% \tablewidth{0pt}
% \label{tab:parameters_and_priors}
% \tablehead{ Parameter & Prior range & WEAKF & ME1+1 & ME2 & NOGR1+1 & NOGR2 & LINGR1+1 & LINGR2 }
% \startdata
% $\Bpar$ & $[-3000,3000]$ G & 1 \\
% $\Bperp$ & $[0,4000]$ G & 1 \\
% $\Delta \lambda_\mathrm{dopp}$ & $[0.01,0.08]$ m\AA & & 2 & 2\\
% $v_\mathrm{LOS}$ & $[-5,5]$ km s$^{-1}$ & & 2 & 2 & 2 & 2 & 4 & 4\\ 
% $\beta$ & $[0,40]$ & & 2 & 2 \\
% $\eta_0$ & $[0,40]$ & & 4 & 4 \\
% $a$ & $[0,0.5]$ & & 2 & 2 \\
% $B$ & $[0,3000]$ G & & 1 & 2 & 1 & 2 & 2 & 4 \\
% $\theta_B$ & $[0,180]$ deg & & 1 & 2 & 1 & 2 & 1 & 2 \\ 
% $\phi_B$ & $[0,180]$ deg & 1 & 1 & 2 & 1 & 2 & 1 & 2 \\ 
% $T$ & $[-3000,3000]$ K & & & & 6 & 6 & 6 & 6 \\ 
% $v_\mathrm{mic}$ & $[0.5,4]$ km s$^{-1}$ & & & & 4 & 4 & 4 & 4 \\ 
% $v_\mathrm{mac}$ & $[0.1,4]$ km s$^{-1}$ & & & & 1 & 1 & 1 & 1 \\ 
% $f$ & $[0,1]$ & & 1 & 1 & 1 & 1 & 1 & 1 \\
% \cutinhead{}
% Total number of parameters & & 3 & 16 & 19 & 17 & 20 & 20 & 24
% \enddata
% \label{tab:parameters_and_priors}
% \end{deluxetable*}
%%%%%%%%%%%%%%%%%%%%%%%%%%%%%

\begin{table*}[!b]
\caption{Model parameters and prior ranges.}
\centering
\scriptsize
\begin{tabular}{ccccccccc}
Parameter & Prior range & WEAKF & ME1+1 & ME2 & NOGR1+1 & NOGR2 & LINGR1+1 & LINGR2 \\
\hline \hline
$\Bpar$ & $[-3000,3000]$ G & 1 \\
$\Bperp$ & $[0,4000]$ G & 1 \\
$\Delta \lambda_\mathrm{dopp}$ & $[0.01,0.08]$ m\AA & 1 & 2 & 2\\
$v_\mathrm{LOS}$ & $[-5,5]$ km s$^{-1}$ & 1 & 2 & 2 & 2 & 2 & 4 & 4\\ 
$\beta$ & $[0,40]$ & & 2 & 2 \\
$\eta_0$ & $[0,40]$ & & 4 & 4 \\
$a$ & $[0,0.5]$ & 1 & 2 & 2 \\
$B$ & $[0,3000]$ G & & 1 & 2 & 1 & 2 & 2 & 4 \\
$\theta_B$ & $[0,180]$ deg & & 1 & 2 & 1 & 2 & 1 & 2 \\ 
$\phi_B$ & $[0,180]$ deg & 1 & 1 & 2 & 1 & 2 & 1 & 2 \\ 
$T$ & $[-4000,4000]$ K & & & & 6 & 6 & 6 & 6 \\ 
$v_\mathrm{mic}$ & $[0,4]$ km s$^{-1}$ & & & & 4 & 4 & 4 & 4 \\ 
$v_\mathrm{mac}$ & $[0.1,4]$ km s$^{-1}$ & & & & 1 & 1 & 1 & 1 \\
$d$ & $[0,1]$ & 1 & & & & & & \\
$f$ & $[0,1]$ & & 1 & 1 & 1 & 1 & 1 & 1 \\
\hline
Total number of parameters & & 7 & 16 & 19 & 17 & 20 & 20 & 24
\end{tabular}
\label{tab:parameters_and_priors}
\end{table*}

\begin{table*}
\caption{Log-evidences for all profiles under different models.}
\centering
\scriptsize
\begin{tabular}{c|ccccccccccc}
 & $V$ shape & SNR$_\mathrm{V}$ & SNR$_\mathrm{L}$ & WEAKF & ME1+1 & 
ME2 & NOGR1+1 & NOGR2 & LINGR1+1 & LINGR2 \\
\hline \hline
Class 0 & Network & 23.42 & 2.77 &    1723.81  &    1660.62  &    1753.04  &    1809.28  &    1860.67  &  \textbf{\textcolor{red}{  1906.46}}  &    1869.03
\\  
Class 1 & Blue-lobe & 19.35 & 4.08 &    1681.35  &    1749.58  &    1823.81  &    1832.26  &    1784.11  &  \textbf{\textcolor{red}{  1869.85}}  &    1821.19
\\  
Class 2 & Asymm. & 8.61 & 2.69 &  \textit{\textcolor{red}{  2012.98}}  &    1935.33  &  \textbf{\textcolor{red}{  2013.65}}  &    1988.22  &    1985.49  & 
  1948.34  &    1958.16 \\  
Class 4 & Network & 121.61 & 11.34 &     376.78  &    -176.43  &    1488.50  &    1135.08  &    1750.46  &    1347.22  &  \textbf{\textcolor{red}{  1870.46}}
\\  
Class 9 & Antisymm. & 7.02 & 3.36 &  \textit{\textcolor{red}{  2020.51}}  &    1931.57  &    2005.87  &  \textbf{\textcolor{red}{  2023.87}}  &    2011.58
 &    1968.85  &    1968.04 \\  
Class 11 & Asymm. & 14.90 & 8.39 &    1866.99  &    1784.53  &    1989.53  &    1987.54  &  \textbf{\textcolor{red}{  2005.74}}  &    1867.97  &    1948.63
\\  
Class 17 & Red-lobe & 7.92 & 3.46 &    1799.50  &    1917.24  &  \textbf{\textcolor{red}{  2020.91}}  &    2010.74  &    1997.67  &    1986.40  &    1970.23
\\  
Class 25 & $Q$-like & 6.38 & 3.27 &    1891.65  &    1986.42  &    1962.10  &  \textbf{\textcolor{red}{  2003.31}}  &    1987.00  &    1983.41  &    1976.43
\\  
Class 34 & $Q$-like & 6.81 & 7.16 &    1846.99  &    1942.06  &  \textbf{\textcolor{red}{  1974.09}}  &    1966.59  &    1947.35  &    1938.07  &    1912.06
\\  
Penumbra & \nodata & 130.91 & 71.10 &  -16732.98  &     336.36  &  \textbf{\textcolor{red}{  1096.64}}  &    -342.23  &     540.81  &    -312.53  & 
   796.47 \\  
Umbra & \nodata & 25.61 & 10.79 &    -125.80  &    1286.08  &  \textbf{\textcolor{red}{  1350.64}}  &    1137.25  &    1290.49  &    1136.49  &    1271.89 \\
IMaX1 & Large $V$ & 2.83 & 4.52 &  \textbf{\textcolor{red}{    50.73}}  &      41.30  &      20.84  &     -34.56  &     -57.58  &     -41.33  &    -105.31 \\

IMaX2 & Large $QU$ & 9.80 & 1.70 &  \textbf{\textcolor{red}{    30.42}}  &      20.99  &  \textit{\textcolor{red}{    29.90}}  &     -53.33  &     -49.52  & 
   -42.52  &     -94.61 \\  
IMaX3 & Large $QUV$ & 6.14 & 5.10 &      12.67  &  \textbf{\textcolor{red}{    27.55}}  &      13.25  &     -39.50  &     -61.06  &     -49.78  &     -82.14
\\  
IMaX4 & Weak $QUV$ & 1.30 & 1.15 &  \textbf{\textcolor{red}{    69.83}}  &      26.97  &       4.93  &     -30.29  &     -81.86  &     -63.25  &    -140.28
\end{tabular}
\label{tab:evidences}
\end{table*}

\begin{table*}
\caption{Bayesian Information Criterion for all profiles under different models.}
\centering
\scriptsize
\begin{tabular}{ccccccccc}
& WEAKF & ME1+1 & ME2 & NOGR1+1 & NOGR2 & LINGR1+1 & LINGR2 \\
\hline \hline
Class 0 &   1349.17  &   1473.11  &   1272.03  &   1047.49  &    951.23  &  \textbf{\textcolor{red}{  845.89}}  &    897.67 \\  
Class 1 &   1436.13  &   1239.75  &   1116.04  &   1003.70  &   1078.10  &  \textbf{\textcolor{red}{  874.41}}  &    944.51 \\  
Class 2 &    771.29  &    889.01  &    745.76  &    683.87  &  \textbf{\textcolor{blue}{  669.06}}  &    755.23  &    717.76 \\  
Class 4 &   4037.33  &   5135.94  &   1767.55  &   2423.49  &   1150.73  &   1978.35  &  \textbf{\textcolor{red}{  832.77}} \\  
Class 9 &    757.79  &    895.62  &    758.04  &  \textbf{\textcolor{red}{  627.73}}  &    645.41  &    717.80  &    703.27 \\  
Class 11 &   1057.44  &   1175.08  &    779.94  &    674.40  &  \textbf{\textcolor{red}{  639.60}}  &    907.18  &    729.46 \\  
Class 17 &   1196.41  &    954.41  &    730.28  &    661.41  &  \textbf{\textcolor{blue}{  647.05}}  &    697.42  &    684.23 \\  
Class 25 &   1014.89  &    806.24  &    803.24  &  \textbf{\textcolor{blue}{  687.35}}  &    706.02  &    717.65  &    705.45 \\  
Class 34 &   1096.30  &    880.69  &    779.09  &  \textbf{\textcolor{blue}{  756.46}}  &    766.41  &    801.63  &    807.33 \\
Penumbra &  38124.65  &   3972.00  &  \textbf{\textcolor{red}{ 2428.08}}  &   5229.84  &   3425.94  &   5118.68  &   2887.95 \\  
Umbra &   3691.00  &    843.90  &  \textbf{\textcolor{red}{  695.22}}  &   1067.04  &    723.85  &   1051.54  &    737.57 \\  
IMaX1 &  \textbf{\textcolor{red}{   77.15}}  &    135.19  &    144.92  &   1521.15  &   1284.42  &   1237.18  &   2160.23 \\  
IMaX2 &  \textbf{\textcolor{red}{  116.64}}  &    182.23  &    130.70  &   2344.45  &    515.75  &   1125.61  &   1414.07 \\  
IMaX3 &  \textbf{\textcolor{blue}{  149.54}}  &    153.08  &    159.19  &   1748.04  &   1788.02  &   1400.30  &   1509.17 \\  
IMaX4 &  \textbf{\textcolor{red}{   35.86}}  &    164.37  &    182.56  &   1650.46  &   2080.83  &   1924.08  &   3796.05
\end{tabular}
\label{tab:bic}
\end{table*}

\begin{table*}
\caption{Properties of the observed Stokes profiles.}
\centering
\scriptsize
\begin{tabular}{ccccccccc}
& $P_\mathrm{tot}$ [\%] & $P_V$ [\%]& $a$ [\%]& $A$ [\%]& $v_V$ [km s$^{-1}$] & $v_I$ [km s$^{-1}$] \\
\hline \hline
Class 0 &  $    2.55$ & $    2.55$ & $   12.48$ & $   39.87$ &  $    0.24$ & $    0.03$ \\ 
Class 1 &  $    2.12$ & $    2.10$ & $   48.98$ & $   49.06$ &  $   -1.65$ & $    0.04$ \\ 
Class 2 &  $    0.94$ & $    0.94$ & $   10.91$ & $  -11.28$ &  $   -1.27$ & $   -0.86$ \\ 
Class 4 &  $   13.23$ & $   13.22$ & $    1.27$ & $   12.84$ &  $   -0.52$ & $   -0.53$ \\ 
Class 9 &  $    0.77$ & $    0.76$ & $   10.51$ & $   14.67$ &  $    0.09$ & $   -0.23$ \\ 
Class 11 &  $    1.82$ & $    1.62$ & $   -0.89$ & $    5.93$ &  $   -0.71$ & $   -0.26$ \\ 
Class 17 &  $    0.95$ & $    0.86$ & $  -17.41$ & $  -35.69$ &  $   -1.72$ & $   -0.27$ \\ 
Class 25 &  $    0.70$ & $    0.69$ & $  -18.42$ & $   43.70$ &  $    4.81$ & $    0.48$ \\ 
Class 34 &  $    0.86$ & $    0.74$ & $  -$ & $  -$ &  $   -4.27$ & $   -1.06$ \\ 
Penumbra &  $   18.93$ & $   17.40$ & $    4.50$ & $   -2.84$ &  $   -0.17$ & $   -0.05$ \\ 
Umbra &  $   21.10$ & $   21.08$ & $   -4.49$ & $   -5.19$ &  $   -0.17$ & $    2.13$ \\ 
IMaX1 &  $    0.53$ & $    0.28$ & $-$ &  $-$ &  $   -3.24$ & $    1.08$ \\ 
IMaX2 &  $    0.99$ & $    0.98$ & $-$ &  $-$ &  $    2.36$ & $    1.69$ \\ 
IMaX3 &  $    0.70$ & $    0.61$ & $-$ &  $-$ &  $   -0.43$ & $   -0.00$ \\ 
IMaX4 &  $    0.17$ & $    0.13$ & $-$ &  $-$ &  $   -3.17$ & $    1.31$
\end{tabular}
\label{tab:properties_profiles}
\end{table*}

\begin{table*}
\caption{Values of the reduced $\chi^2$ for the best fits and their ratios.}
\centering
\scriptsize
\begin{tabular}{cccccccccccc}
& $\chi^2_\mathrm{ME1+1}$ & $\chi^2_\mathrm{ME2}$ & $\frac{\chi^2_\mathrm{ME2}}{\chi^2_\mathrm{ME1+1}}$ & 
$\chi^2_\mathrm{NOGR1+1}$ & $\chi^2_\mathrm{NOGR2}$ & $\frac{\chi^2_\mathrm{NOGR2}}{\chi^2_\mathrm{NOGR1+1}}$ & 
$\chi^2_\mathrm{LINGR1+1}$ & $\chi^2_\mathrm{LINGR2}$ & $\frac{\chi^2_\mathrm{LINGR2}}{\chi^2_\mathrm{LINGR1+1}}$\\
\hline \hline
Class 0 &      3.07 &      2.58 &      0.84 &      2.11 &      1.85 &      0.88 &      1.62 &      1.68 &      1.04 \\
Class 1 &      2.55 &      2.23 &      0.88 &      2.01 &      2.13 &      1.06 &      1.68 &      1.78 &      1.06 \\
Class 2 &      1.77 &      1.41 &      0.80 &      1.29 &      1.22 &      0.94 &      1.41 &      1.28 &      0.90 \\
Class 4 &     11.25 &      3.69 &      0.33 &      5.18 &      2.30 &      0.44 &      4.14 &      1.53 &      0.37 \\
Class 9 &      1.78 &      1.43 &      0.80 &      1.17 &      1.17 &      1.00 &      1.33 &      1.24 &      0.93 \\
Class 11 &      2.40 &      1.48 &      0.62 &      1.27 &      1.16 &      0.91 &      1.75 &      1.30 &      0.74 \\
Class 17 &      1.91 &      1.37 &      0.72 &      1.24 &      1.17 &      0.94 &      1.28 &      1.20 &      0.93 \\
Class 25 &      1.58 &      1.53 &      0.97 &      1.30 &      1.30 &      1.00 &      1.33 &      1.25 &      0.94 \\
Class 34 &      1.75 &      1.48 &      0.85 &      1.46 &      1.44 &      0.99 &      1.52 &      1.48 &      0.97 \\
Penumbra &      8.65 &      5.16 &      0.60 &     11.44 &      7.37 &      0.64 &     11.15 &      6.12 &      0.55 \\
Umbra &      1.67 &      1.29 &      0.78 &      2.15 &      1.34 &      0.62 &      2.07 &      1.32 &      0.64
\end{tabular}
\label{tab:chi2_table}
\end{table*}

\subsection{Milne-Eddington models}
The second simplest model we consider is based on the Unno-Rachkovsky solution of the radiative transfer equation 
in a Milne-Eddington atmosphere \citep[see][for details]{harvey72,auer_heasly_house77,landi_landolfi04}.
Under this approximation, we assume that the ratio between the line absorption coefficient
and the continuum absorption coefficient does not vary with depth in the atmosphere. The
same happens with the bulk velocity of the plasma and the magnetic field vector.
Additionally, we assume that the source function varies linearly with optical
depth along the LOS. Each (magnetic or non-magnetic) component is characterized by 
a vector of physical quantities $\thetabold$ that contains: 
the Doppler width of the line in wavelength units ($\Delta \lambda_\mathrm{dopp}$), the wavelength 
shift due to a macroscopic bulk velocity ($v_\mathrm{LOS}$),
the gradient of the source function ($\beta$), the ratio between the line and continuum absorption coefficients ($\eta_0$) for
each line and a line damping parameter ($a$). This vector is augmented in the magnetized components
with the magnetic field vector parameterized by its modulus, inclination and azimuth
with respect to the local vertical direction ($B$, $\theta_B$ and $\phi_B$, respectively). 
Additionally, a filling factor ($f$) is included to weight the two components following Eq. (\ref{eq:filling_factor}). 
The specific equations used in our code are shown in App. A.
The number of unknowns is 16 for the case of one field-free component plus a magnetic one (labeled
ME1+1) and 19 for the case of two magnetic components (labeled ME2). Note that the number of 
wavelength points of the IMaX data is similar to the number of free parameters. Consequently, we
expect the model selection scheme to favor simpler models (with less number of free parameters) for IMaX observations.

We use uniform priors for all variables, with the ranges indicated in Table \ref{tab:parameters_and_priors}.
It is important to note that the ranges of the parameters have to be set up realistically since
they affect the final value of the evidence. The reason is that the evidence is the integral
of the normalized likelihood weighted by the normalized prior. As a consequence, the larger the prior volume, the smaller
the evidence. We consider that the values shown in Table \ref{tab:parameters_and_priors} are
a good representation of what we expect a priori. In any case, we have empirically
tested that modifying the range of the parameters (always making sure not to cut
regions of the space of parameters that are compatible with the observations) has a 
small effect on the evidence because of the dominance of the integral of the likelihood. However, adding a new parameter strongly
modifies the evidence because of the increased dimensionality of the space of parameters. Note that we avoid the 180$^\circ$ ambiguity
by restricting the azimuth to lay between 0$^\circ$ and 180$^\circ$.

The computation of the evidence is carried out with \B\ \citep{asensio_martinez_rubino07}\footnote{All codes can be
freely downloaded from the webpage \texttt{http://www.iac.es/proyecto/magnetism}.}, the Bayesian
inference code for Milne-Eddington atmospheres. This code makes use of the 
Multinest algorithm \citep[][]{feroz09}, based on the nested sampling approach
introduced by \cite{skilling04}. One of the free parameters of Multinest is the number of
live points \citep[see][for more details]{feroz09}, which is directly related to the final precision of the estimate of the
evidence. We have verified that $n_\mathrm{live}=600$ gives enough precision for our purposes.
Concerning the computing time, each inference with \B\ takes of the order of one
minute, being slightly dependent on the number of free parameters.

% \subsection{MISMA Milne-Eddington models}
% Since asymmetries are of importance in many of the analyzed profiles, we follow 
% \citep{misma96,sanchez_almeida_misma96} and propose a simplified model that
% takes into accound micro-structured atmospheres. This approximation is based
% on three Milne-Eddington components. One of them can have the presence of 
% a magnetic field or not and produces output Stokes profiles $\mathbf{S}_1$. 
% Of the remaining two (which are always magnetized), one of them is a standard 
% Milne-Eddington atmosphere characterized by the vector of parameters defined previously.
% The other one shares the parameters $\beta$ and $\eta_0$ but can have different
% values of the rest of parameters. Both components are combined so that the total
% propagation matrix is given by the weighted addition of propagation matrices, 
% following the multicomponent approach of \citep{sanchez_almeida_misma96} 
% to mimetize the microscopic averaging of the propagation matrix:
% \begin{equation}
% \mathbf{K} = \alpha \mathbf{K}_2 + (1-\alpha) \mathbf{K}_3,
% \end{equation}
% where $\alpha \in [0,1]$. This produces an emerging
% Stokes vector $\mathbf{S}_2$ after applying the analytical solution of the radiative
% transfere equation in the Milne-Eddington approximation. Finally, the emergent Stokes
% profiles are obtained as:
% \begin{equation}
% \mathbf{S} = f \mathbf{S}_1 + (1-f) \mathbf{S}_2,
% \end{equation}
% where $f \in [0,1]$.

\subsection{Models with gradients along the LOS}
Some of the Stokes profiles analyzed in this work present asymmetries, both in area
and in amplitude. It is known 
that amplitude asymmetries can be produced with the presence of more than one
magnetic component (even with Milne-Eddington atmospheres), something that 
we already take into account in Eq. (\ref{eq:filling_factor}). On the contrary, area
asymmetries can only be produced under the
presence of correlated gradients between the velocity and the
magnetic field vector along the LOS. For this reason
we also consider models that acommodate such
gradients and are in local thermodynamical equilibrium (LTE), which constitute our
most complex models.

Following previous approaches \citep{sir92,socas_navarro98,frutiger00}, the
Harvard-Smithsonian Reference Atmosphere model \citep[HSRA;][]{gingerich71} is used as
a starting point. In this model, the physical quantities defined in Table 
\ref{tab:parameters_and_priors} are perturbed (within the range presented in
the table) at predefined positions 
(usually termed nodes) to improve the fitting. A polynomial function in the
continuum optical depth is fit to the value of the nodes 
and added directly to the original HSRA stratification. The order of the
polynomial depends on the number of nodes considered for each
physical quantity. If only one node is chosen, the HSRA stratification
is perturbed by adding a constant at every height but keeping the original gradients. If two
nodes are chosen, a straight line is used to modify the HSRA stratification, thus
introducing additional gradients. In the case of three nodes, a parabola modifies the
gradient and the curvature. For instance, a parabolic function is added to 
the HSRA temperature stratification, where the value at three nodes in the
atmosphere are chosen in the range $[-4000,4000]$ K.

\begin{figure*}
\centering
\includegraphics[width=0.43\textwidth]{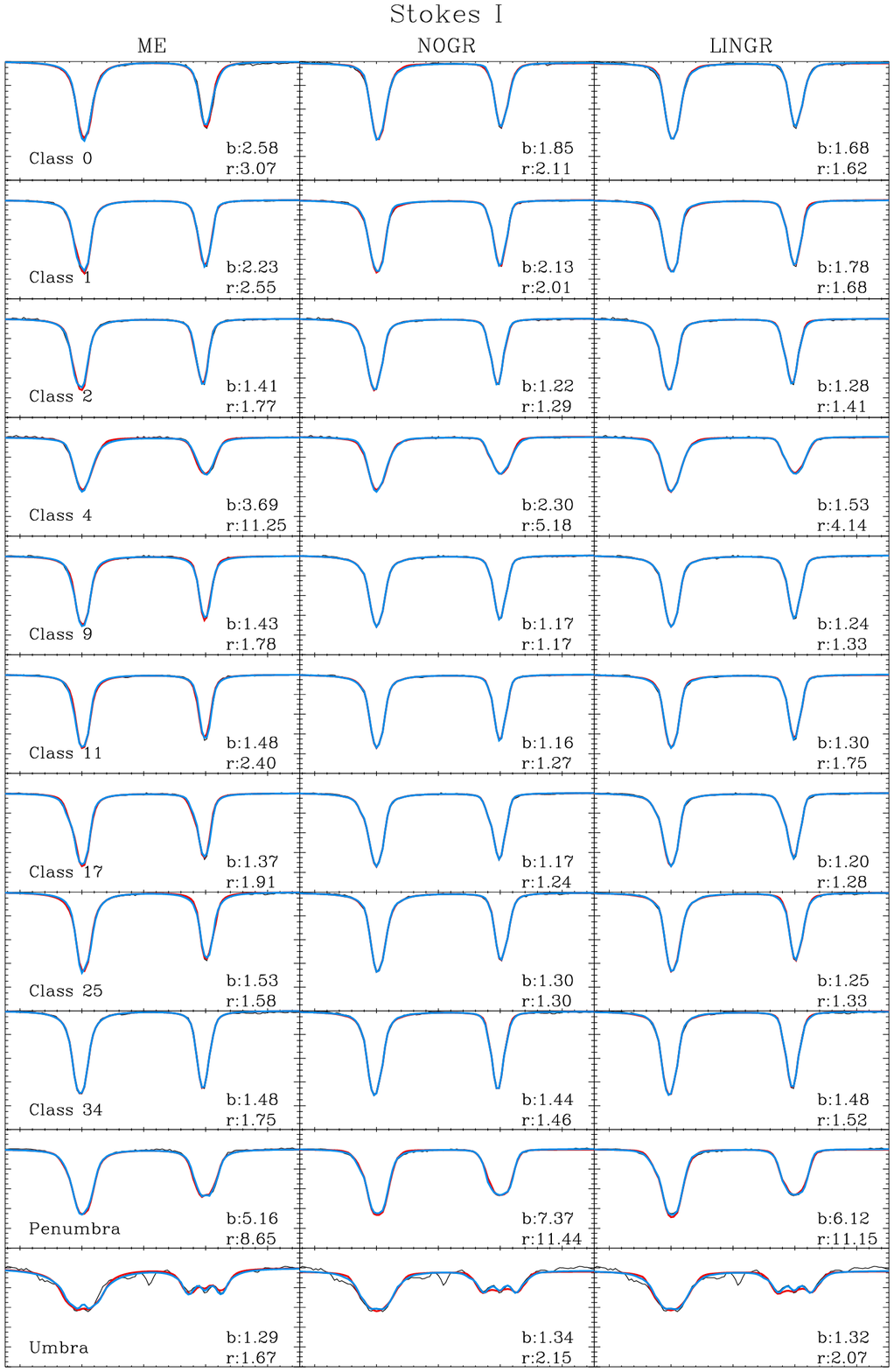}
\includegraphics[width=0.43\textwidth]{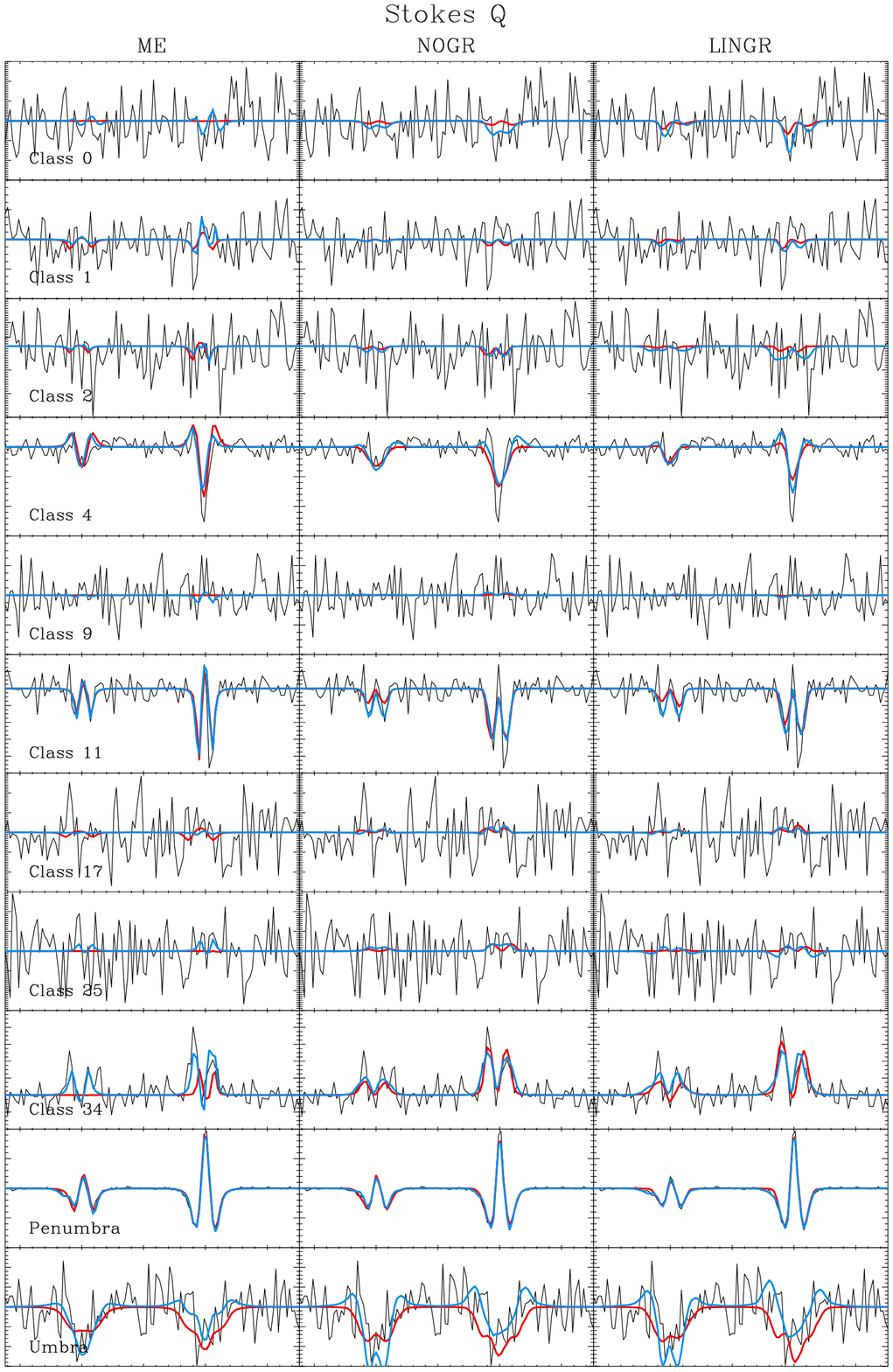}\vspace{0.2cm}
\includegraphics[width=0.43\textwidth]{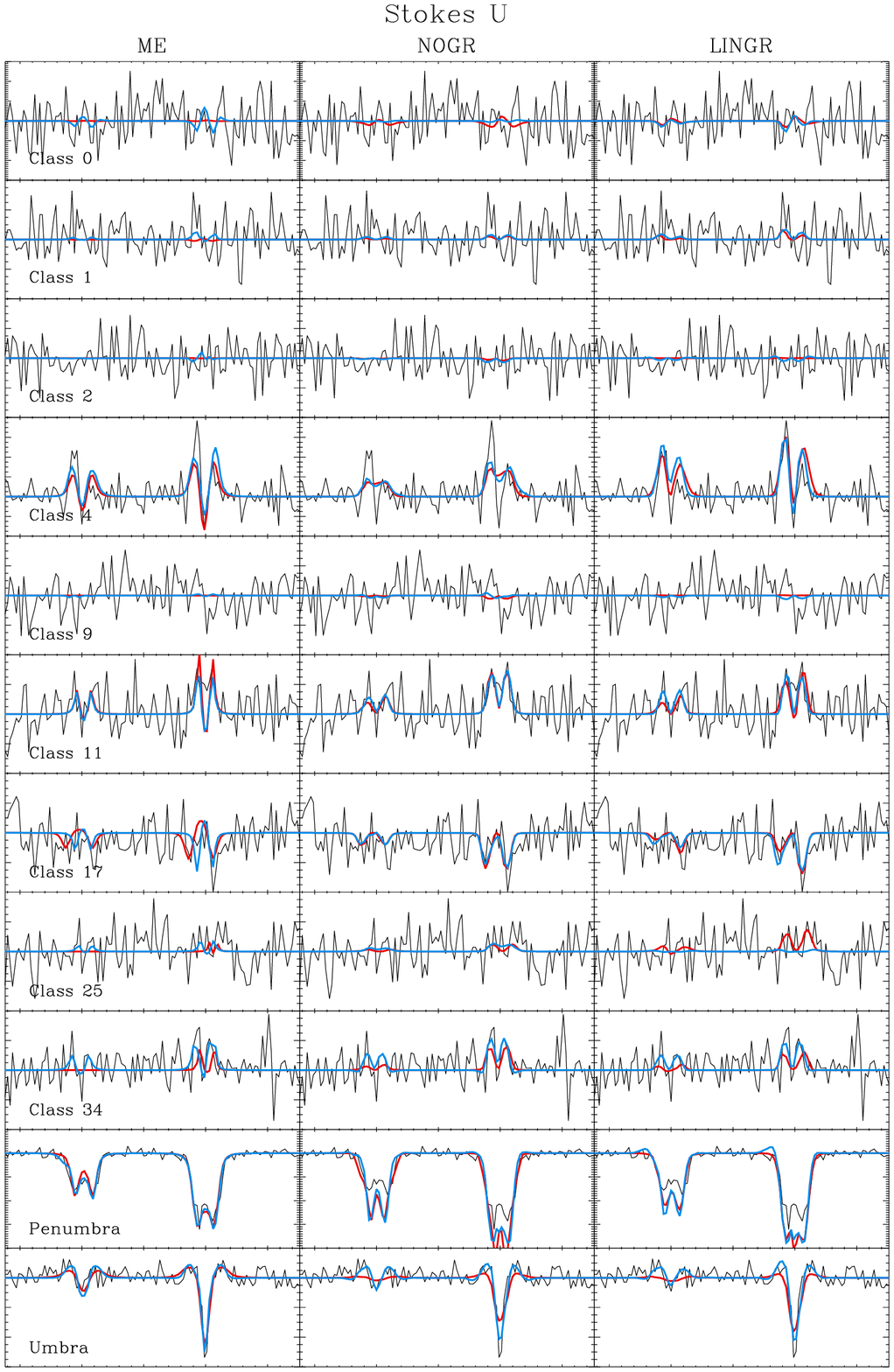}
\includegraphics[width=0.43\textwidth]{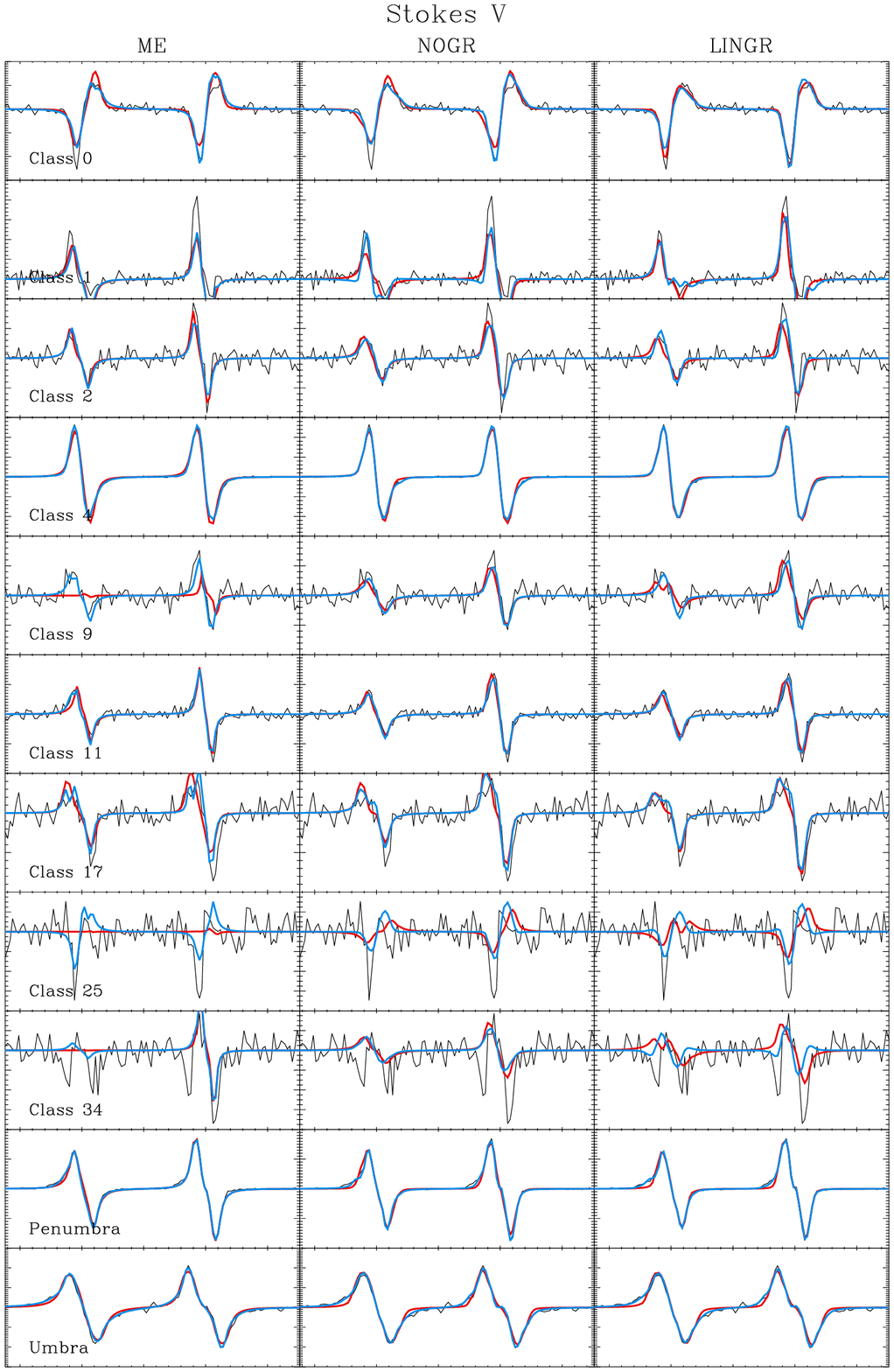}
\caption{Maximum a-posteriori fits to the observed Stokes profiles. The black curves are the observed
Stokes profiles. The red curves correspond to the models with one magnetic and one non-magnetic components, while
the the models with two magnetic components correspond to the blue profiles. We indicate the value of the
$\chi^2$ for each fit, where the labels $b$ and $r$ refer to ``blue'' and ``red'' curves, respectively.}
\label{fig:map_fits}
\end{figure*}

The number of nodes selected for each physical parameter is displayed 
on Table \ref{tab:parameters_and_priors}. We fix the number of nodes for
the temperature to three, so that the total amount of nodes is six for the 
two components.
In order to
test which is the importance of gradients, we consider two options for the LOS velocity
and the magnetic field strength. The first one assumes that both quantities
take a constant value throughout the atmosphere (models NOGR1+1 and NOGR2, depending on the
number of magnetic components). The second one introduces a linear gradient with optical
depth (models LINGR1+1 and LINGR2). This opens up the generation of correlated gradients
which can potentially generate asymmetries in the Stokes profiles. The number of nodes of 
the rest of the parameters is kept fixed in both models.

The evidence is calculated with \BL, an updated version of \B\ which makes use of an
accelerated version of the synthesis core of Nicole (Socas-Navarro, de la Cruz Rodr\'{\i}guez,
Asensio Ramos, Trujillo Bueno \& Ruiz Cobo, in preparation) and it is also based
on the Multinest algorithm. The computation cost of \BL\ is larger than that for
\B\ given the large number of evaluations of the model. Each inference with \BL\ takes of the order of 10-20
minutes, being slightly dependent on the number of free parameters.
The synthesis engine uses the Hermitian formal solver of \cite{bellot98} and obtains the pressure scale by putting the model in hydrostatic equilibrium using a
neural network approach to speed up the calculation. Once synthesized, the lines are convolved with a 
macroturbulent velocity ($v_\mathrm{mac}$) to increase the broadening and produce
a better fitting.

\section{Results and discussion}

\subsection{Maximum a-posteriori profiles}
Although a potentially large space of parameters is compatible with the observations,
the set of parameters that maximize the full multidimensional posterior (maximum a-posteriori; MAP) is of some
interest. Since we use flat priors, this solution is equivalent to the one that maximizes
the likelihood. Additionally, because we use a Gaussian likelihood, this solution is
also equivalent to the one that minimizes the standard $\chi^2$ metric used in standard least squares
inversion codes. It is important to stress that this solution is not distinguished in any special way from 
all those that fit the profiles inside the error bars. 

Fig. \ref{fig:map_fits} shows the observed Hinode Stokes $I$, $Q$, $U$ and $V$ in black, together with the best
fits. The first column corresponds to the fits with Milne-Eddington models, the second
to LTE models without gradients along the LOS on the field strength and velocity and
the third column to LTE models with gradients. In each column, the red curves correspond to the case
of one magnetic component plus a field-free one, while the blue curves 
refer to the case of two magnetic components. Compared with typical inversion results,
one would say that all models are able to do a good job on fitting the full set
of profiles. The value of the reduced $\chi^2$ metric is shown in each panel, for the
1+1 components red (\emph{r}) and 2 components blue (\emph{b}) models. They are
displayed again in Tab. \ref{tab:chi2_table}, where we also show the reduced $\chi^2$ ratio between the
model with 2 components and the model with 1+1 components. We get small
values (close to 1) for all fits, indicating that the fits are acceptable, something that
is also relevant from a pure visual inspection. In general, we find a quite systematic decrease of the
reduced $\chi^2$ when adding a second magnetic component, meaning that the fit is marginally
improved. The decrease of the reduced $\chi^2$ when adding a second magnetic component
is especially large for a few profiles (e.g., class 4 and penumbra). They coincide with the
observations with the largest signal-to-noise (SNR). This is an expected behavior in general, consequence of an increase on the
number of free parameters, that gives a larger flexibility to the model to fit more details
of the observed Stokes profiles. If the SNR of the observation is large, even a relatively good
fit obtained with the 1+1 models will produce a large $\chi^2$ (this is what happens with class 4 and penumbra), which will 
be largely reduced by allowing more freedom to the model (using two magnetic components).
However, this is not always the case, given that some ratios
reported in Tab. \ref{tab:chi2_table} are larger than 1. This is the case for classes 0 and 1
in the models with gradients along the LOS. They correspond to profiles with a low
polarization amplitude and strong Stokes $V$ asymmetries.
In summary, it is hard to say from these fits which model is preferred 
over the others. Pragmatically, in light of the quality of the fits, one would choose
the simplest model. However, the question is whether an improvement in the $\chi^2$ is worth
the increase in the number of parameters.

Model selection is carried out by comparing evidences, i.e., computing the integral of the posterior
over all model parameters. Therefore, the specific values of the parameters is irrelevant. However, for the 
sake of completeness, we have displayed in Tab. \ref{tab:map_table} of Appendix \ref{sec:app_map}
the maximum a-posteriori values of some parameters for all atmospheric models considered and
for all observed Stokes profiles. Concerning the magnetic flux density, we see that the inferred
value is quite robust to the specific model, except for the umbral profile (in which this quantity is not related to
the amplitude of the Stokes $V$ profile) and some IMaX profiles (for which the information is
scarce). This is a consequence of the fact that, if the magnetic field strength is not too strong, the magnetic flux
density is almost an observable. On the contrary, there is a large variability on the inferred value of the filling factor, something that
directly influences the field strength and the inclination. We conclude that the inferred MAP values 
depend on the selected model and that model selection turns out to be important.

\subsection{Model comparison}
\begin{figure*}
\centering
\includegraphics[width=0.3\textwidth]{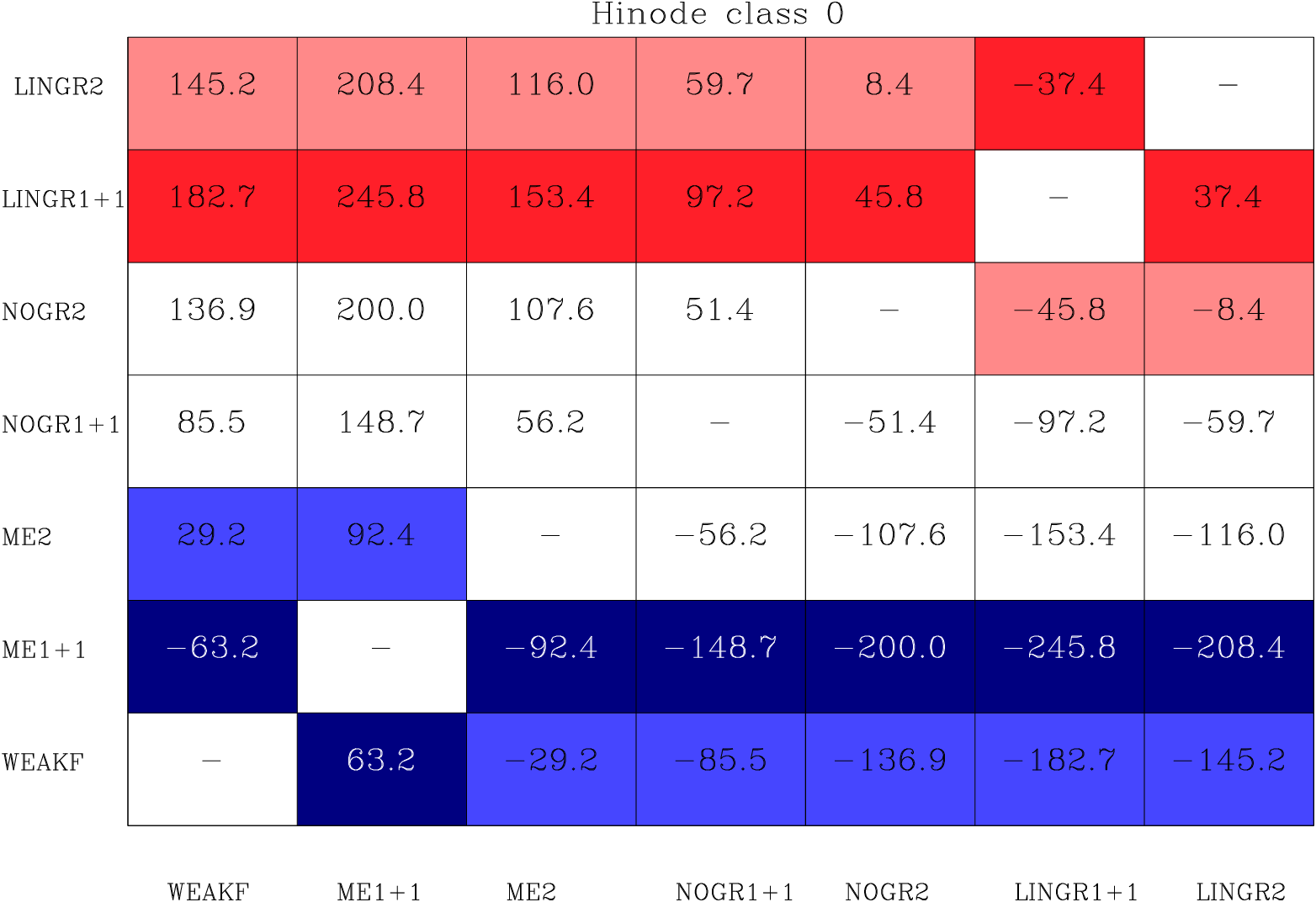}%
\hspace{0.2cm}
\includegraphics[width=0.3\textwidth]{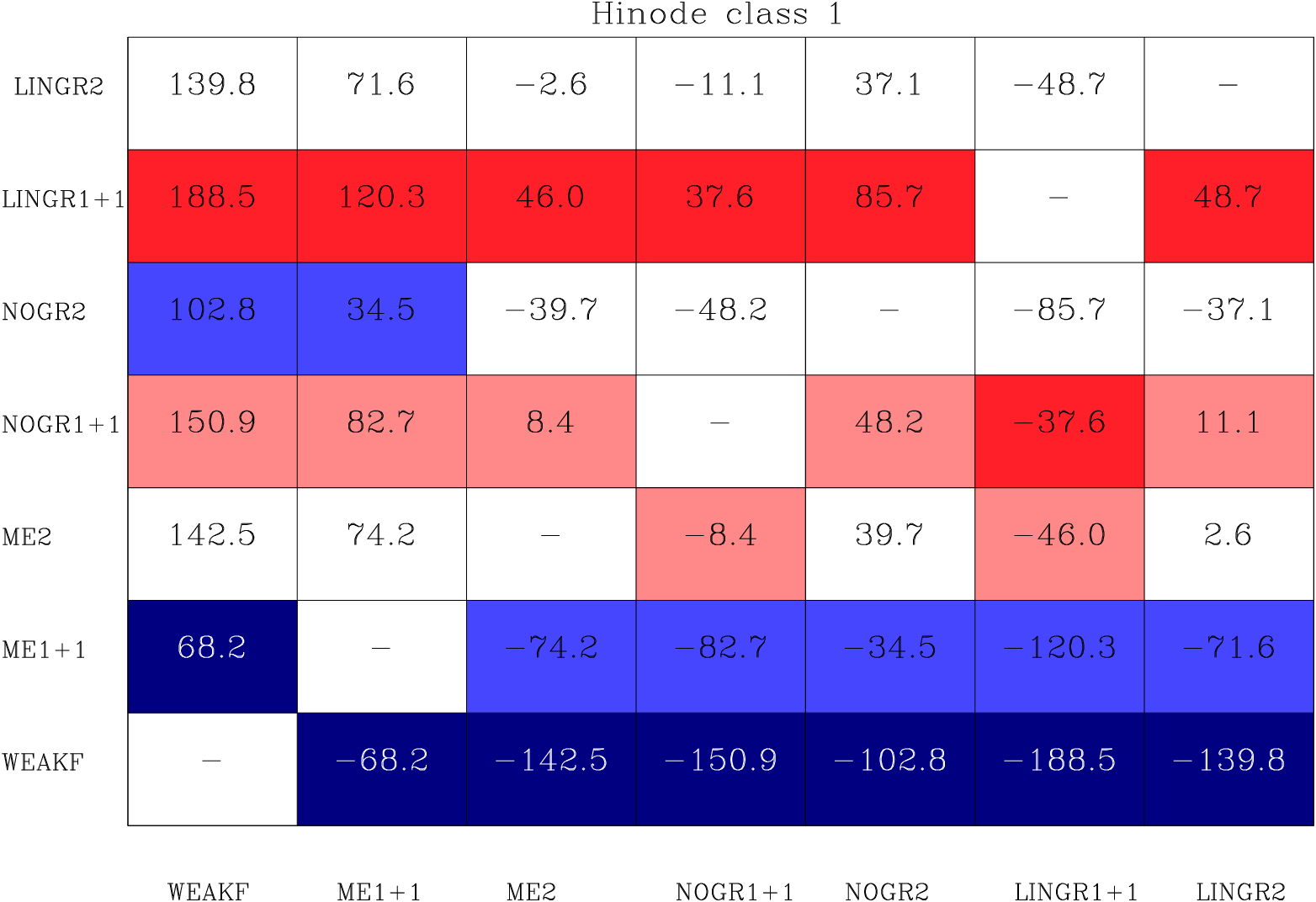}%
\hspace{0.2cm}
\includegraphics[width=0.3\textwidth]{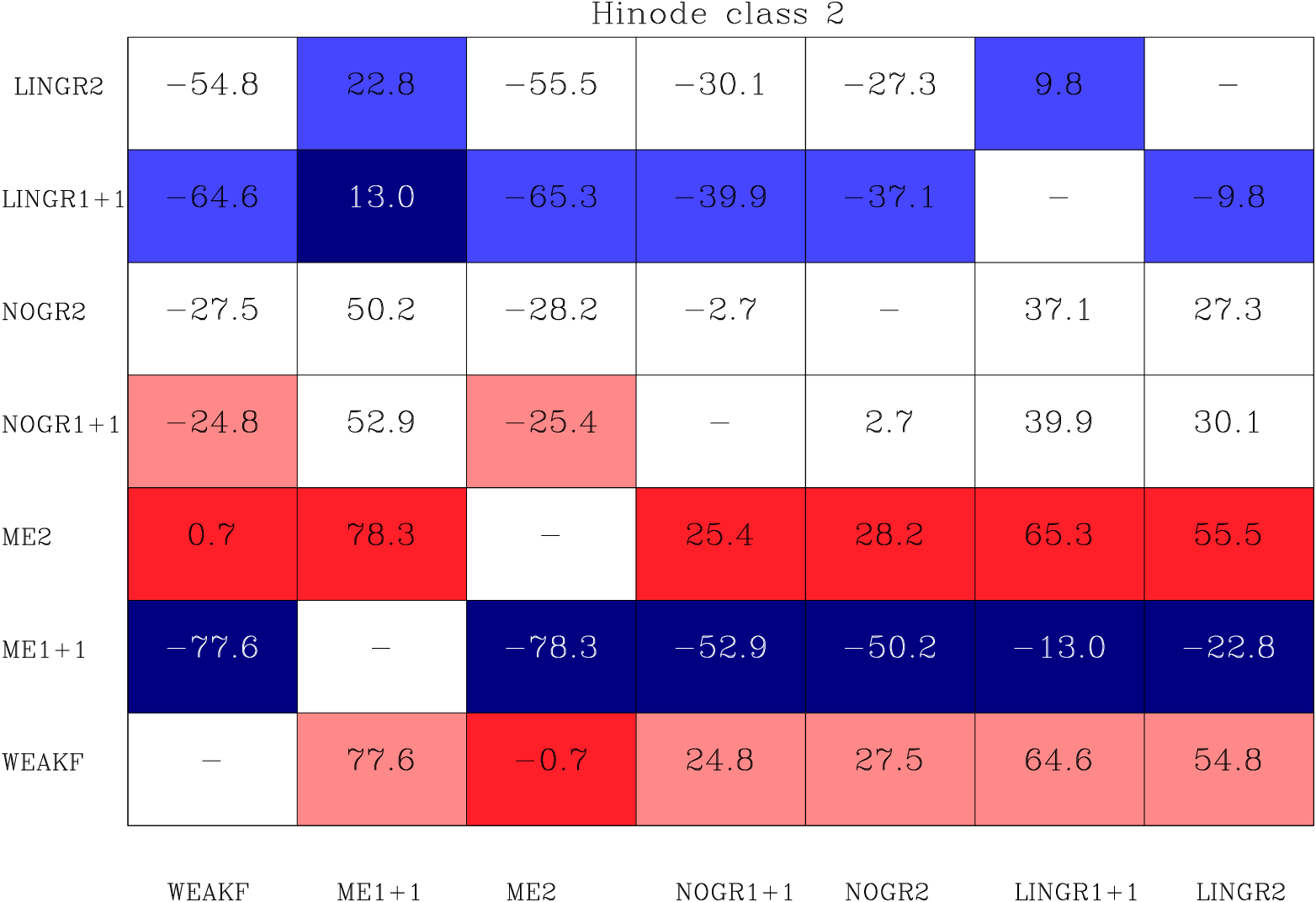}\vspace{0.2cm}
\includegraphics[width=0.3\textwidth]{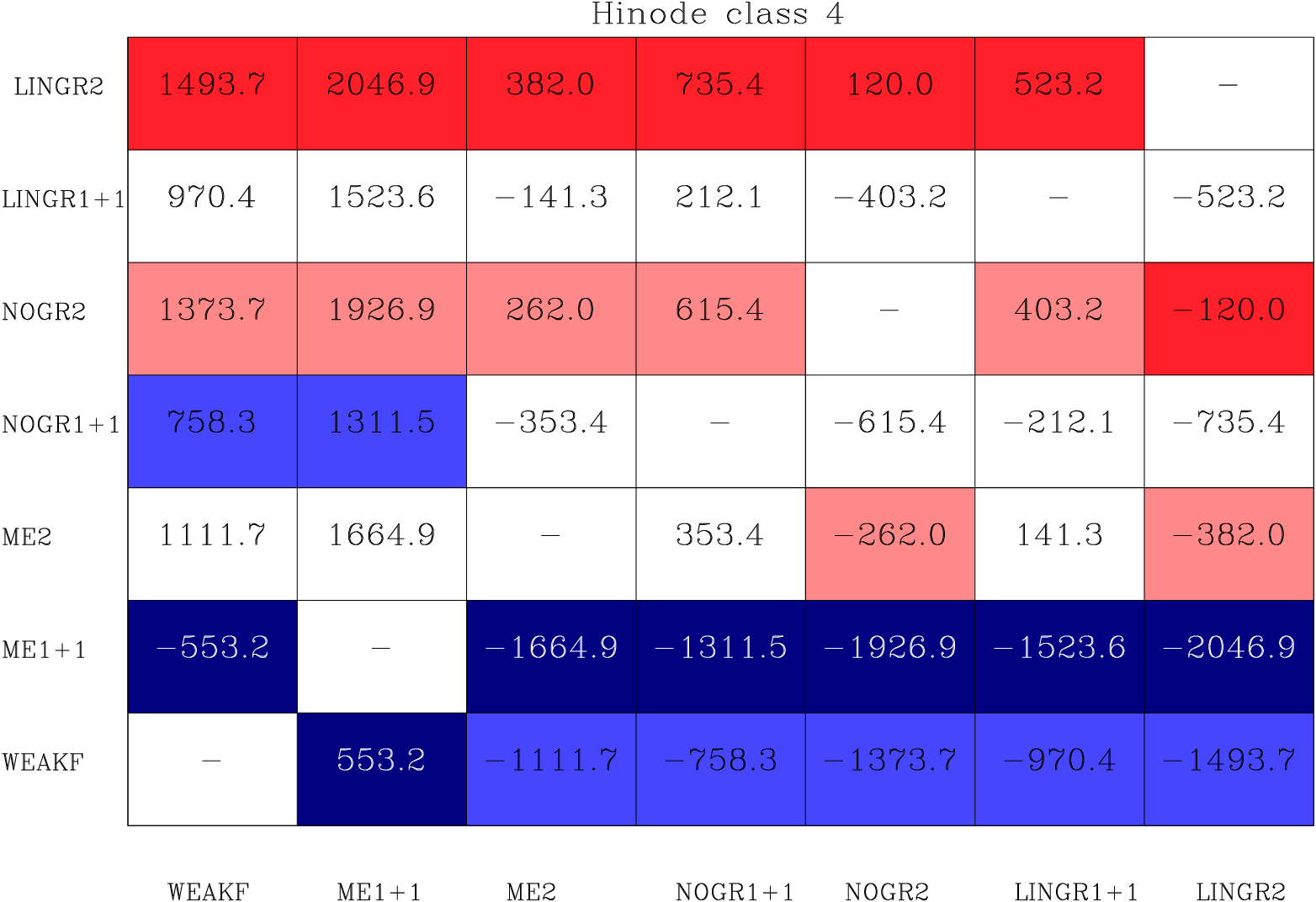}%
\hspace{0.2cm}
\includegraphics[width=0.3\textwidth]{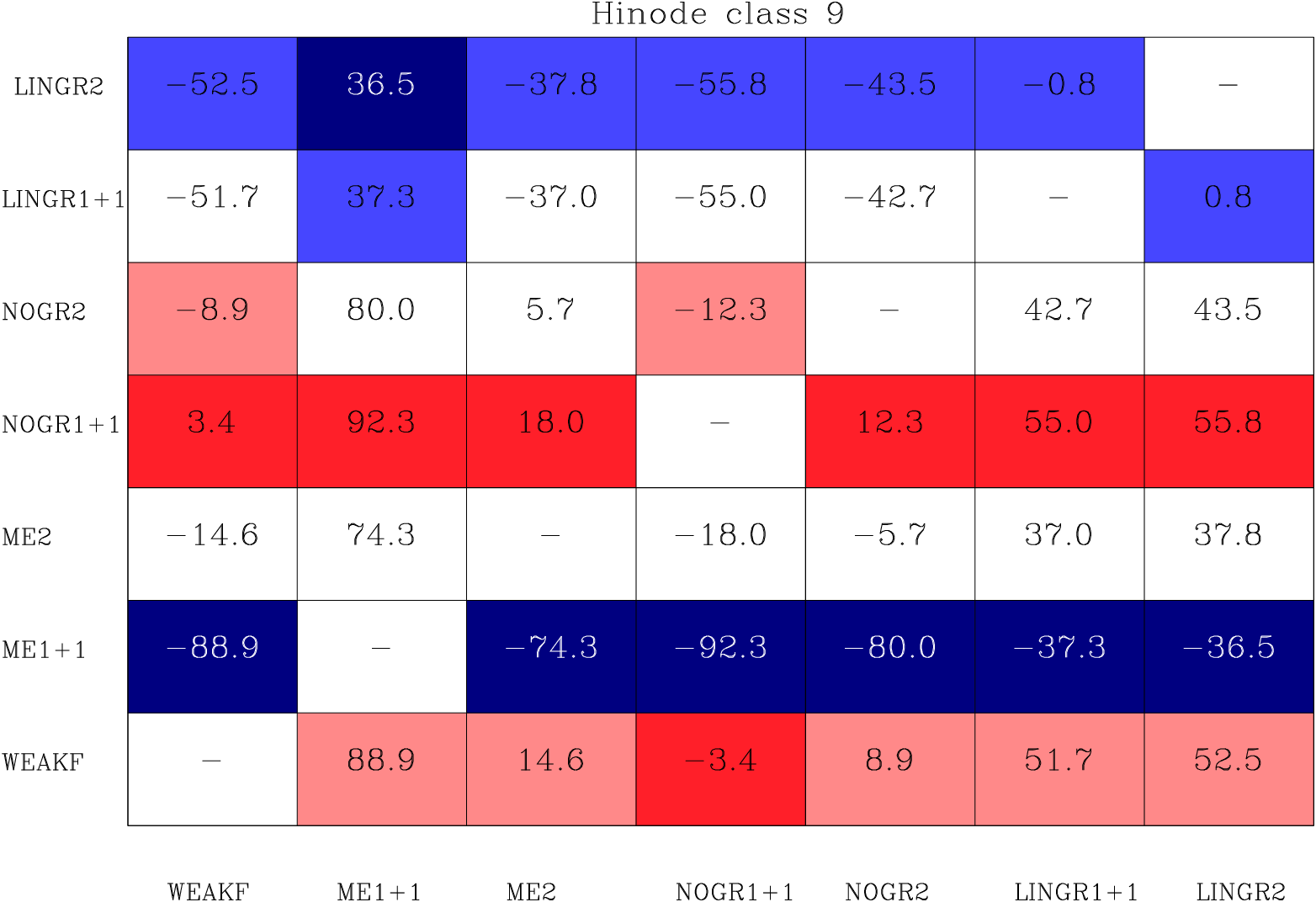}%
\hspace{0.2cm}
\includegraphics[width=0.3\textwidth]{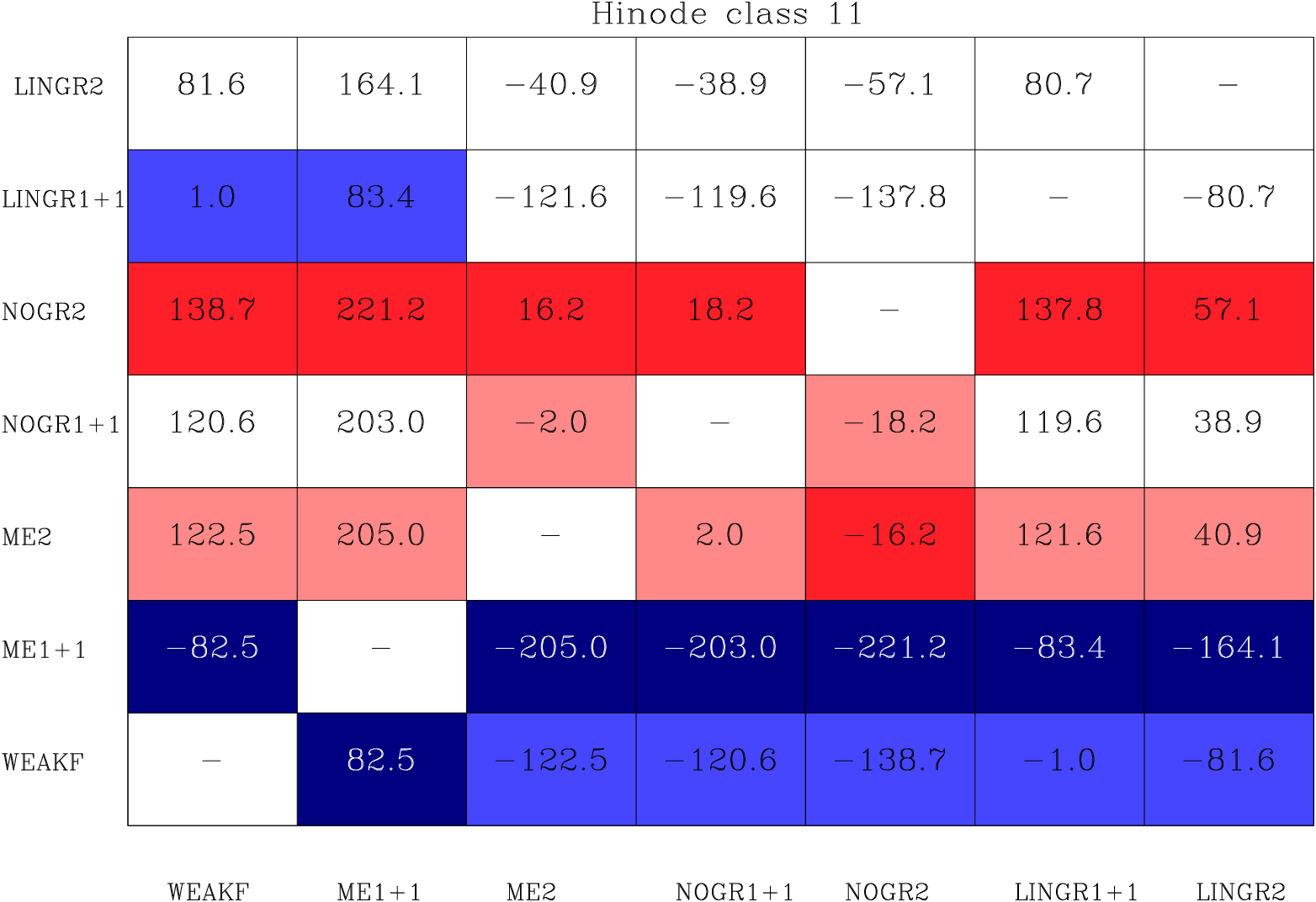}\vspace{0.2cm}
\includegraphics[width=0.3\textwidth]{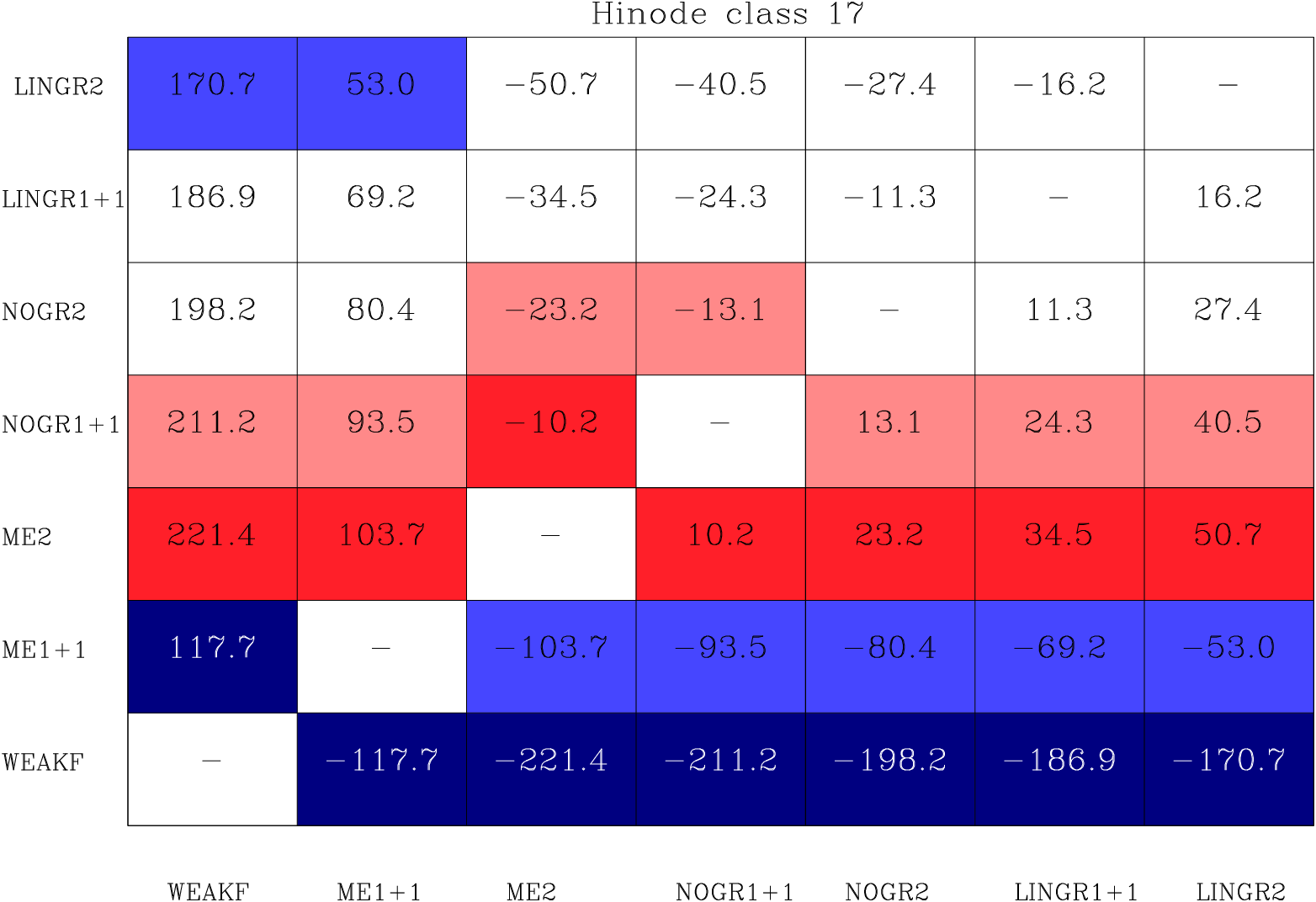}%
\hspace{0.2cm}
\includegraphics[width=0.3\textwidth]{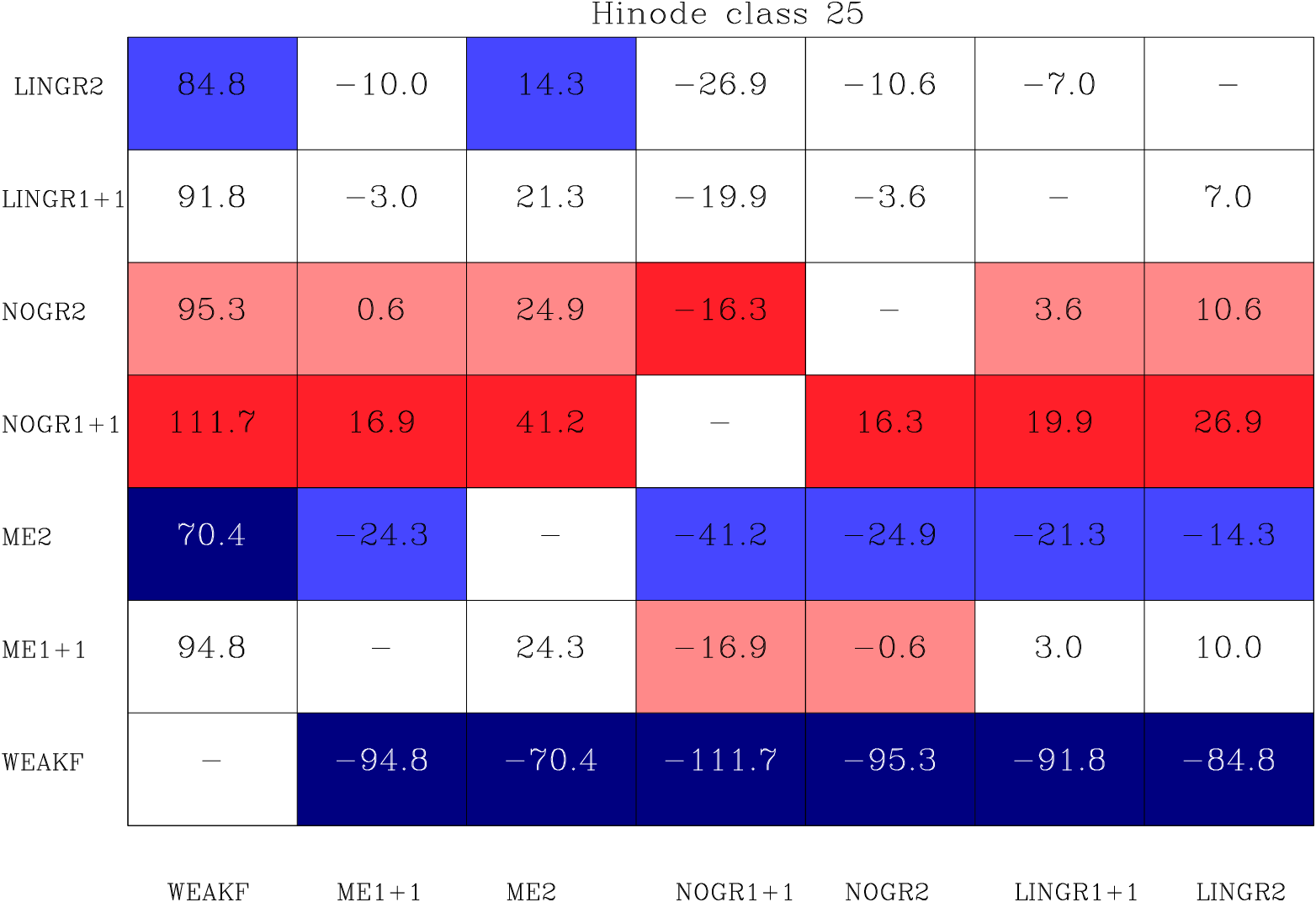}%
\hspace{0.2cm}
\includegraphics[width=0.3\textwidth]{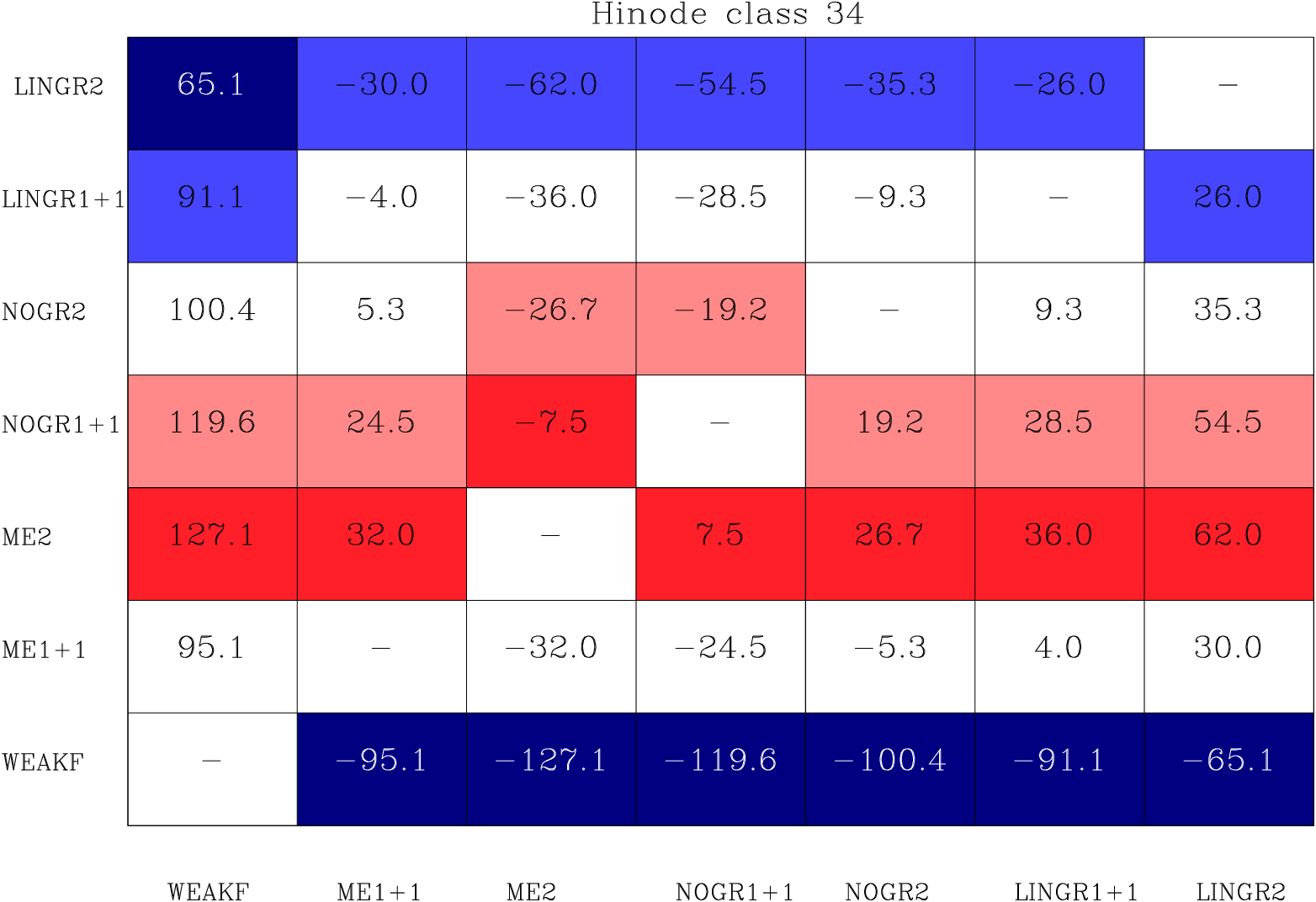}\vspace{0.2cm}
\includegraphics[width=0.3\textwidth]{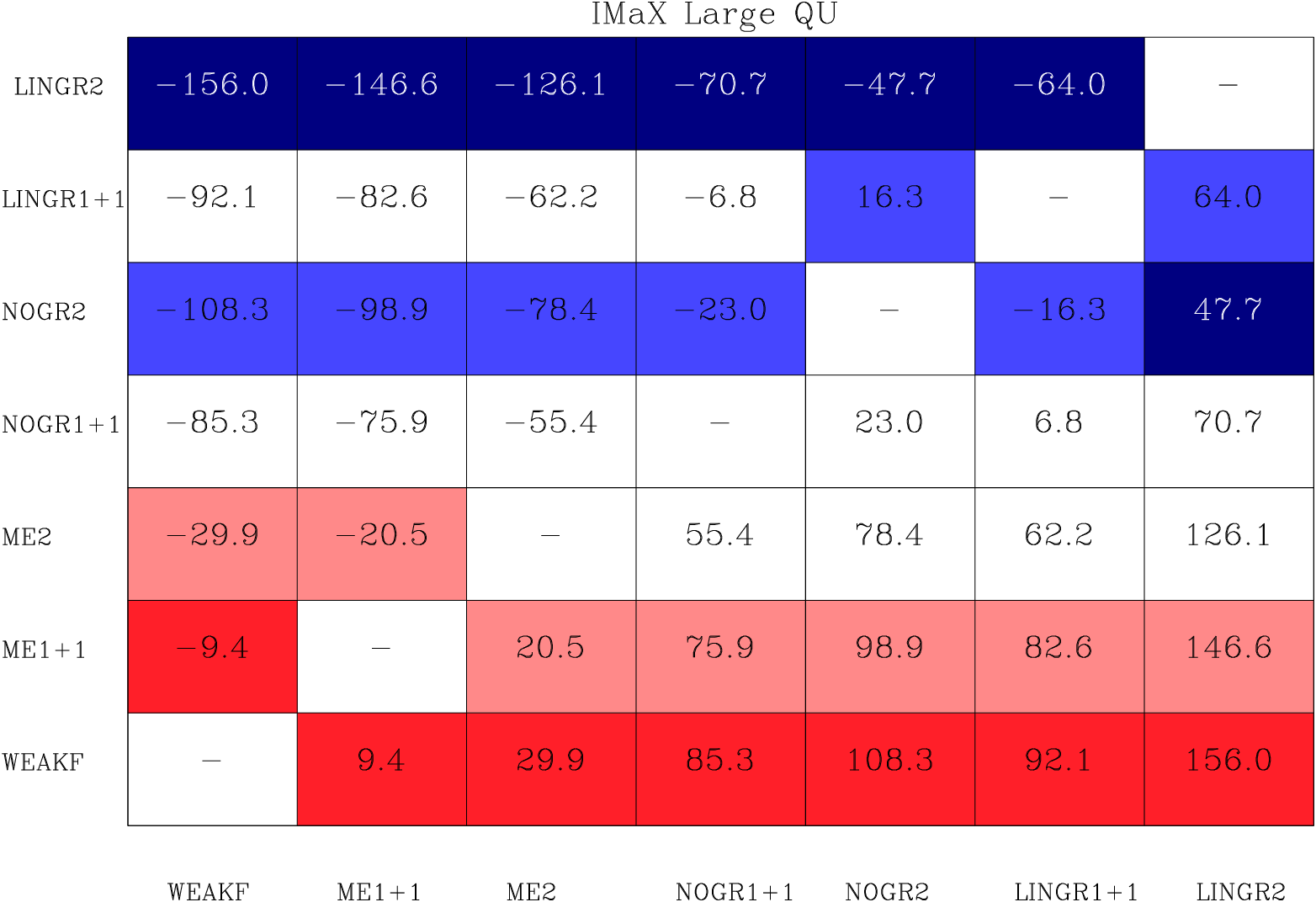}%
\hspace{0.2cm}
\includegraphics[width=0.3\textwidth]{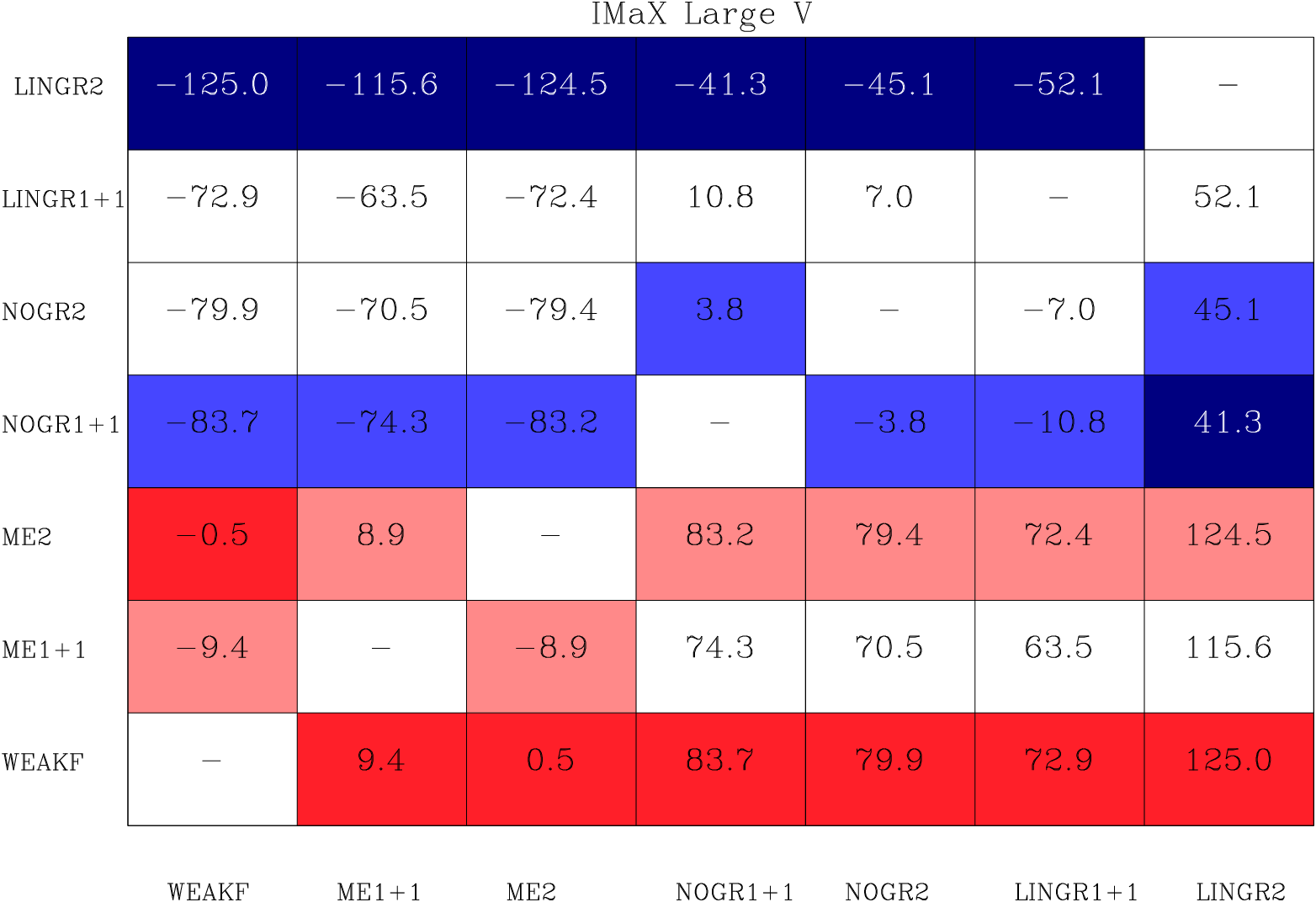}%
\hspace{0.2cm}
\includegraphics[width=0.3\textwidth]{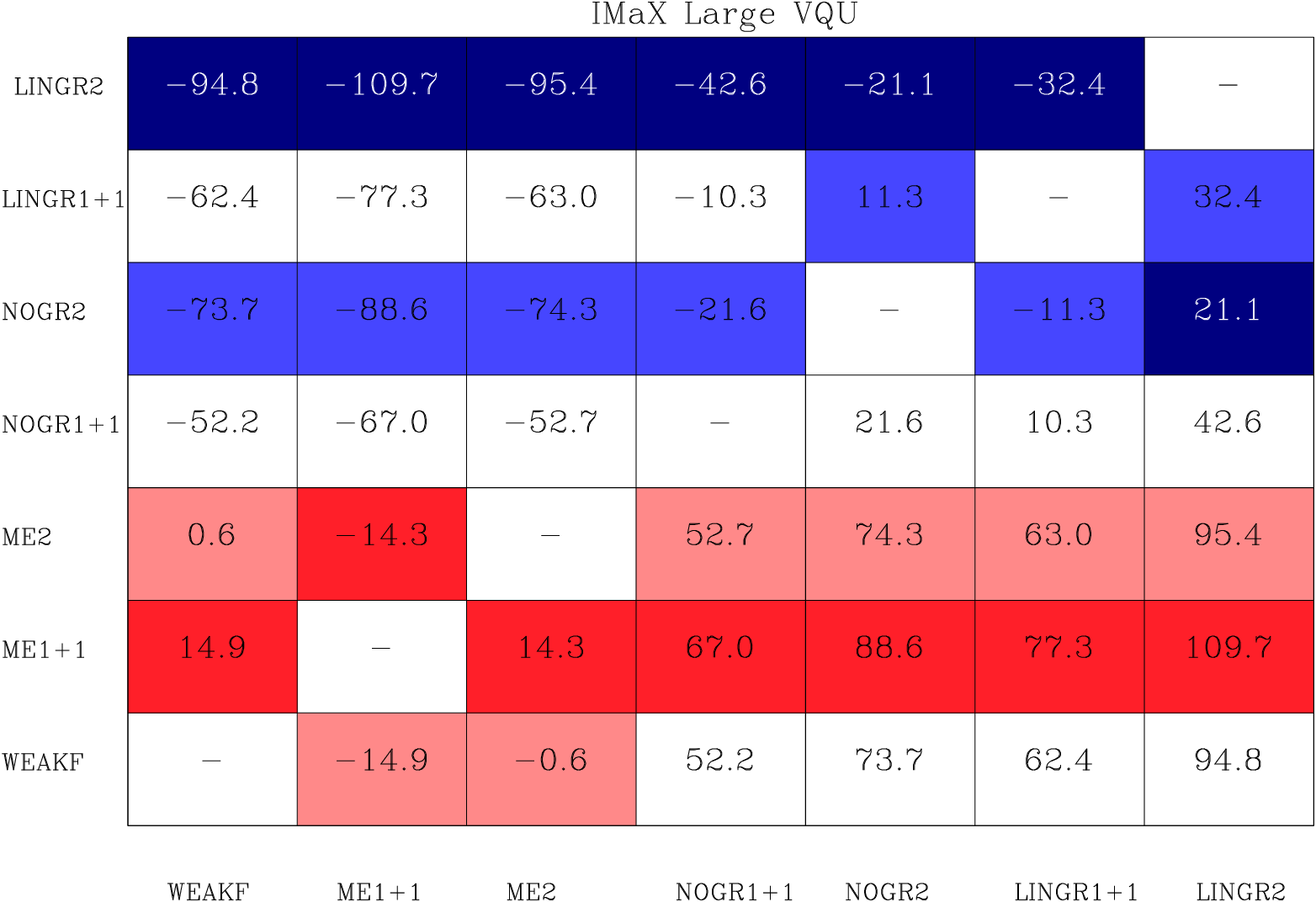}\vspace{0.2cm}
\includegraphics[width=0.3\textwidth]{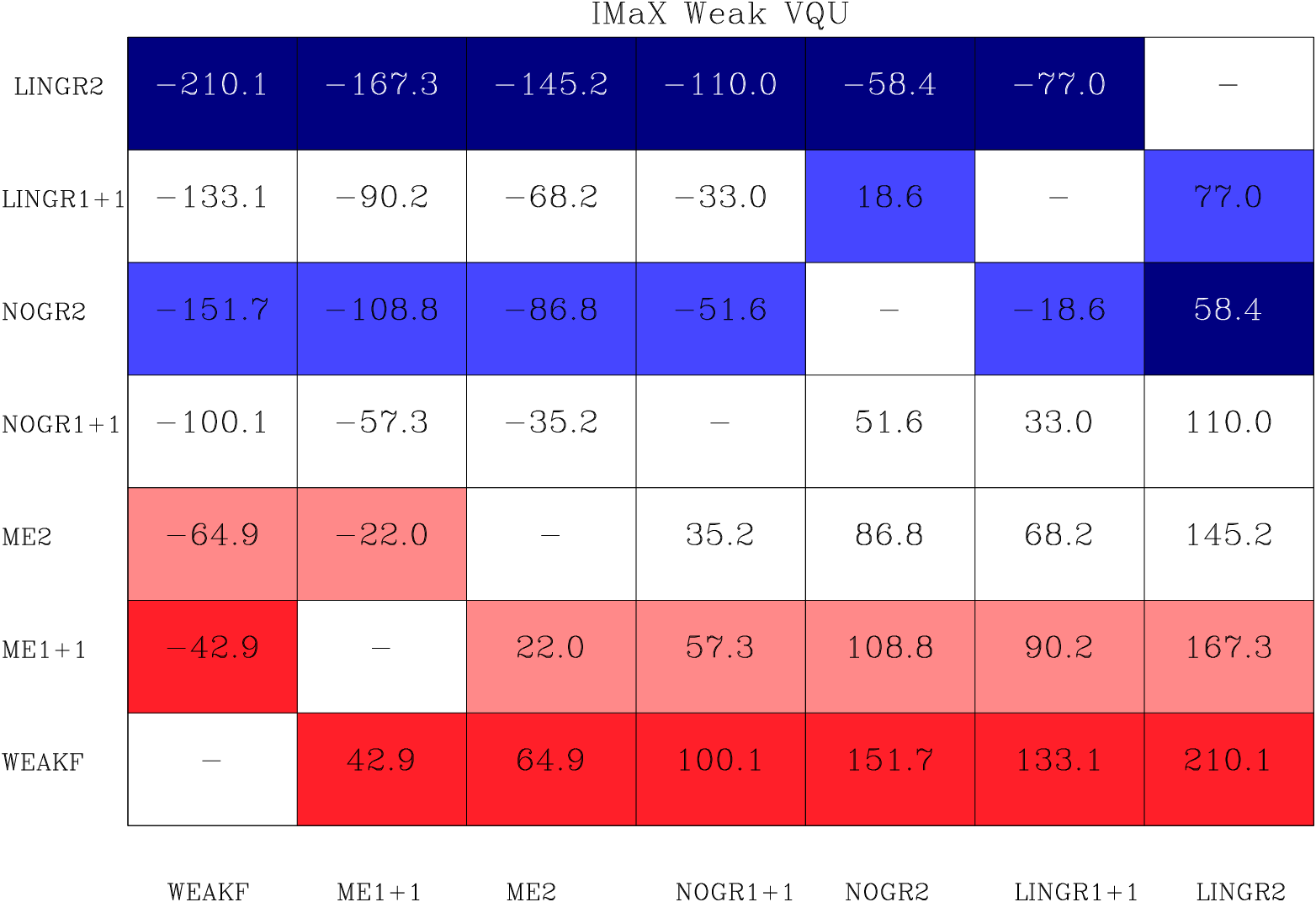}%
\hspace{0.2cm}
\includegraphics[width=0.3\textwidth]{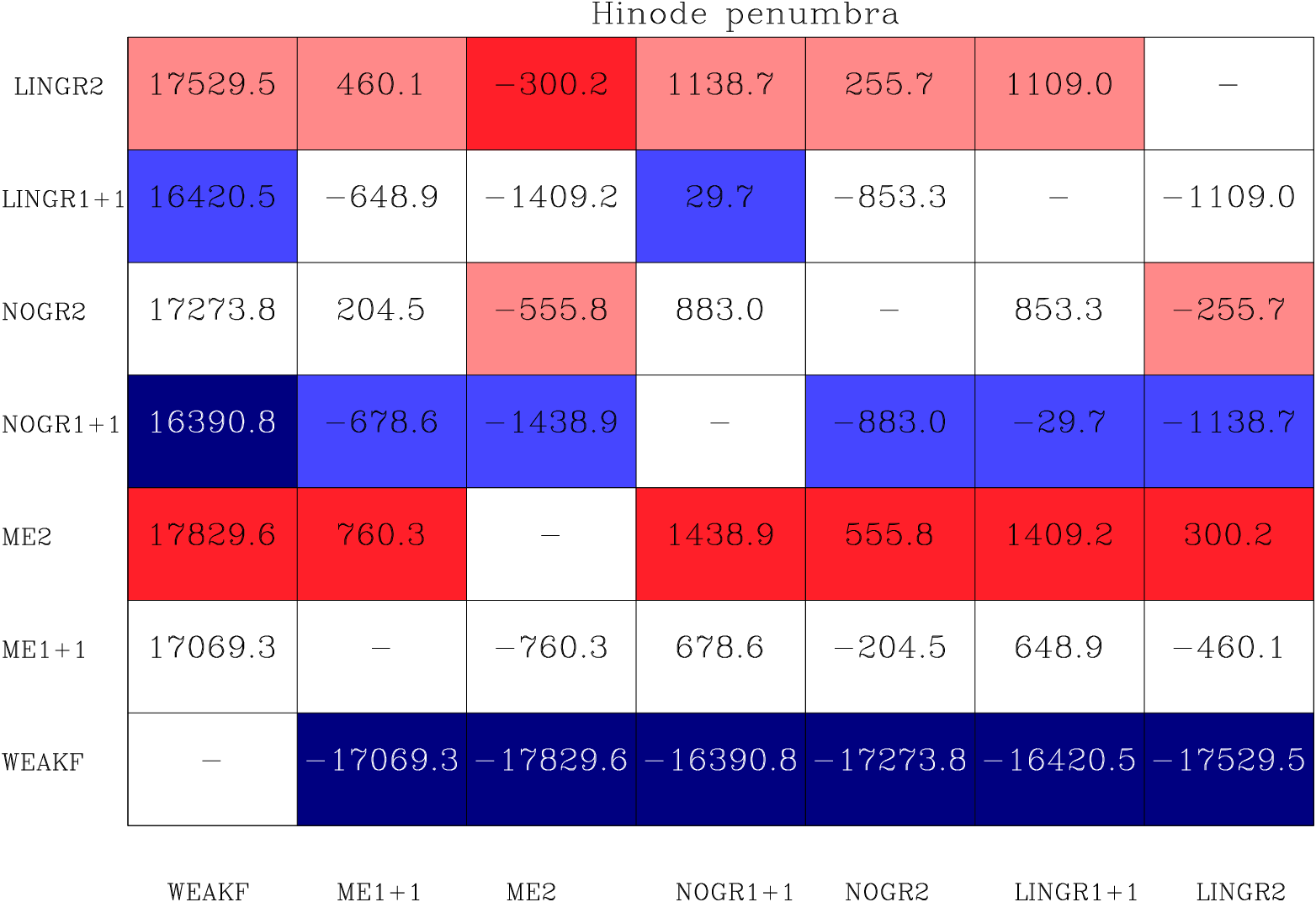}%
\hspace{0.2cm}
\includegraphics[width=0.3\textwidth]{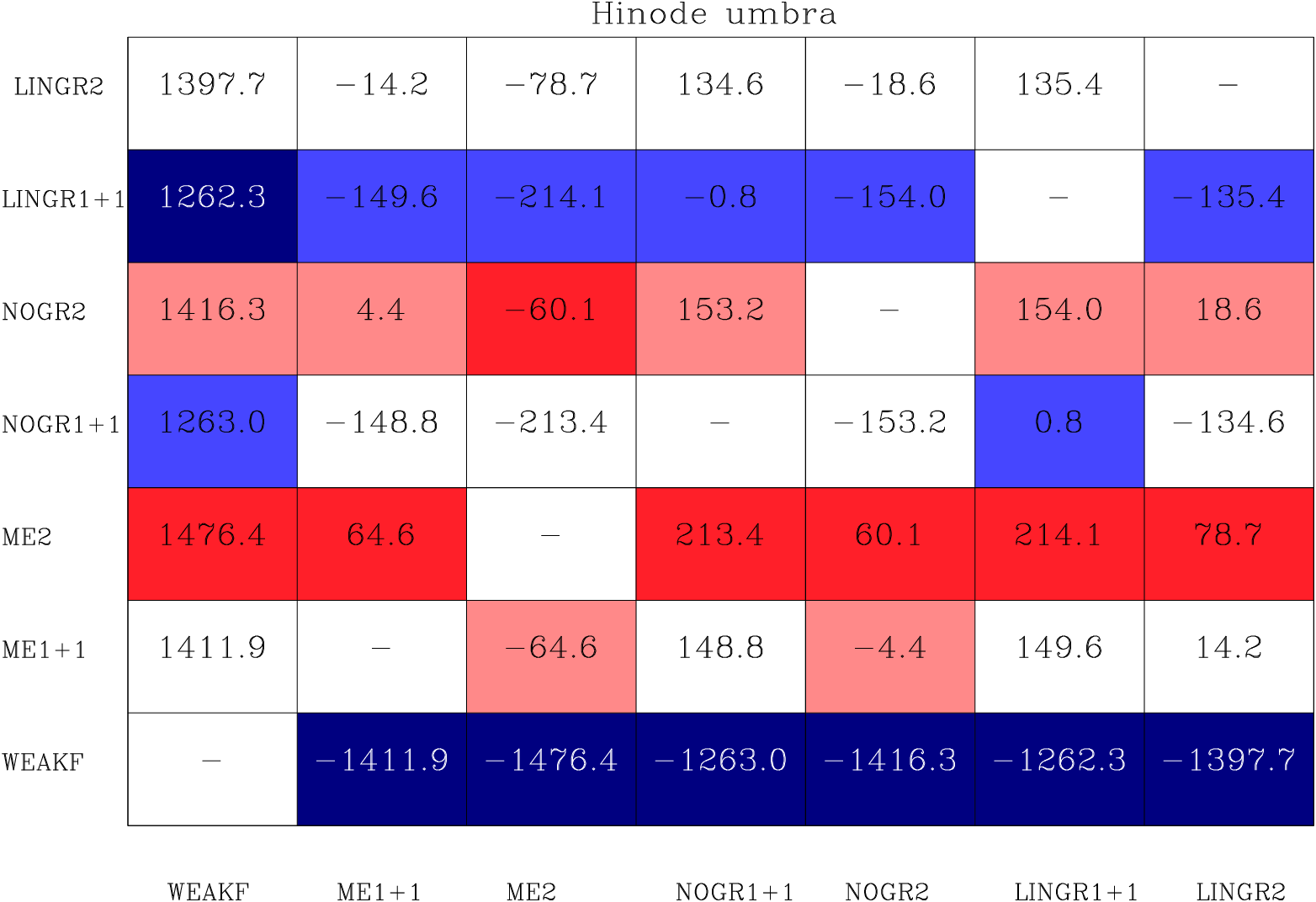}
\caption{Logarithmic evidence ratio from each model with respect to the every other model. Models can be
compared using these tables if we assume the same a-priori probability for all of them. Each square reports
the log evidence ratio between a given model in the vertical axis versus a certain model in the horizontal
axis. Red (and light red) indicate the most probable (and second most probable) model in each column, while blue (and light blue)
show the less probable (and second less probable) model in each column. Note that these tables are symmetric with respect to the diagonal.}
\label{fig:league_evidences}
\end{figure*}

Once the Bayesian evidence is computed for all models considered, model comparison is
just a matter of comparing real numbers and decide on the most probable model following 
Table \ref{tab:jeffreys_scale}. Due to the potentially small/large value of the
evidence, we reported the value of $\ln p(D|\mathcal{M}_i)$ in 
Table \ref{tab:evidences}. We indicate in bold red the model with the largest evidence and
in italic red those models that are in competition with the selected model
according to the Jeffreys' scale.
Given that we cannot discard the presence of more
models compatible with the data, the
evidence cannot be considered as an absolute scale. Consequently, we can only
compute evidence ratios assuming the same a-priori probability for every model.
For this reason, it is also illustrative to consider model comparison in a league
framework. This is shown in Fig. \ref{fig:league_evidences}, where each square
indicates the value of the logarithmic evidence ratio obtained from the competition of
pairs of models (model in the vertical axis versus model in the horizontal axis). Red colors 
point out that the model in the left axis is preferred with respect to the model in the horizontal axis. 
Blue colors are used when the model in the vertical axis is the least preferred one. Light red (blue) colors
are associated with the second most (less) probable model. Obviously, only half of the squares contain
relevant information, with the other half showing redundant data
because they are antisymmetric with respect to the diagonal (with a 
sign change, equivalent to an inverse in a linear scale for the evidence ratio).

The first thing to note is that there is not a single model that can be
considered the ``best'' for all profiles. A Milne-Eddington model
with two magnetic components seems to be the most probable for classes 2, 17 and 34.
According to \cite{viticchie_1_11}, they only represent $\sim$9\% of the
field-of-view. Classes 0, 1 and 4 favor models with gradients along the LOS
in velocity and magnetic field, while classes 9, 11 and 25 favor models without
gradients in velocity and magnetic field. A simple weak-field approximation is
the model of choice for the majority of IMaX data. Moreover, there is not
a single ``worst''  model, although ME1+1 can be considered not appropriate among the
selected models, in general, for explaining our quiet Sun Hinode profiles. This might sound strange
given that many spectropolarimetric inversions of Hinode data have been
carried out with ME1+1 models. Our results demonstrate that this model is often
not preferred by data while the opposite occurs for the ME2 model. The reason
has to be found on the presence of asymmetries that a ME1+1 model cannot fit. Obviously,
this conclusion is based on the limited number of models that we have considered. 
If one proposes a different model (there is no difference whether this is a different atmospheric model or one from
our selection but with some parameters fixed), it is a matter of computing the evidence
to find out if it is preferred by the data.

Another property of
our model comparison is that, normally, one model is orders of magnitude much 
more probable than the rest. Only a few models are really in the position of
competing.  For instance, for class 9, we see that NOGR1+1 has a log-evidence $\sim$3.4
larger than WEAKF. According to Table \ref{tab:jeffreys_scale}, there is moderate
evidence that NOGR1+1 is preferable to WEAKF. On the contrary, for class 2, we can
safely say that there is not a clear preference for ME2 or WEAKF2 (the difference in log-evidence
is smaller than 1), with the two models
being much more probable than the rest. Other examples can be found by looking at the comparisons of Fig. \ref{fig:league_evidences}.

Among the quiet Sun profiles, only for classes 0, 1 and 4 the analysis favors
models with linear gradients on the field strength and LOS velocity, with class 4
preferring a model with two magnetic components. Classes 0 and 4 have been associated 
by \cite{viticchie_1_11} to the network and they present large amplitudes of 
circular and linear polarization (class 0 arriving to 2.55\% and 0.32\% in Stokes $V$
and $L$, respectively, and class 4 reaching 13.2\% and 1.30\% in Stokes $V$
and $L$, respectively). It is apparent from these results that large SNR
ratios are important to favor (or at least allow) more complicated models, especially
when the Stokes profiles are inherently complex (asymmetries, several components clearly
visible). It is important to point out that the higher SNR is the responsible for the need of 
increased complexity just because it does not suppress the important details of the Stokes profiles that
arise in complex atmospheres. For example, classes 0 and 1 favor LINGR1+1 because the SNR
is relatively high and the asymmetry of the Stokes $V$ profiles is clearly visible. Consequently,
an increase in the complexity of the model is compensated by the increase on the quality of
the fit. On the other side we find, for example, class 2, with small asymmetries and favoring simple
models. Additionally, large SNR in Stokes $Q$ and
$U$ are also crucial to favor more complicated models. In our case, class 4 have the largest 
linear polarization signal in the whole set and thus favors complex models even though
the asymmetries of Stokes $V$ are small, while classes 0 and 1 are among the ones with
the smallest amplitude of linear polarization.

In summary, if the SNR is large, it is possible to increase the number of model parameters if we are able to fit
features in the profile that are well above the noise level. This is also consistent
with the fact that the MAP fit gives the smallest $\chi^2$ for class 0, 1 and 4. Note that,
even if the $\chi^2$ of class 0 under LINGR1+1 and LINGR2 models is very similar, LINGR1+1
is preferred because of the smaller number of free parameters.

A ME2 model is preferred for all profiles belonging to classes 2, 17 and 34. In general,
these classes have linear polarization profiles at or below the noise level (except
class 34, which presents clear Stokes $Q$ profiles). Of interest is also the fact that
classes 2 and 9, which have the WEAKF model in competition with more complex models, present
no detection of linear polarization, while the circular polarization profiles are nearly
antisymmetric. The model comparison is suggesting that there is limited information in
such profiles and that a WEAKF model can do a good work extracting all available information.
A different problem is, obviously, how to interpret this information.

A particularly interesting case
is the profile of class 34. The Stokes $V$ signal is clearly $Q$-like as
described by \cite{viticchie_1_11}, giving the idea of several magnetic components
with opposite polarities inside the resolution element. The Bayesian analysis
suggests that a ME2 model is enough. This seems reasonable, although it is
clear that the best fit is not able to correctly fit the two polarities in Stokes $V$.
The reason is that the amplitude of Stokes $Q$ is even larger than that of Stokes $V$
and there is not a clear hint of such two components in linear polarization.

Concerning the Hinode profiles observed in sunspots, Table \ref{tab:evidences} points out
that a ME2 model is the most suitable one. This model already produces a very good 
fit for the two profiles. For the umbra, this preference for ME2 is also understood
at the light of the enhanced noise due to the reduced number of photons or the
presence of molecular lines not accounted for in the inversion.

The comparison of models for explaining IMaX profiles is also very illuminating. The reduced number of
data points strongly suggests that the simple weak-field approximation is the model of
choice for explaining the observations among the ones considered in this work. This is
especially relevant for the profiles in which either circular or linear polarization (or both simultaneously) is small.
This experiment with IMaX data (with the preference for very simple models) shows that 
complex models with a relatively large number of free parameters are only favored when the
number of sampling points in wavelength is large, at least larger than the 25 points
of IMaX data. If the sampling is sufficient, the information encoded in the Stokes
profiles about the model parameters compensate for the increase in the prior volume of complex
models.
In case both polarizations are clearly above the noise level, model ME1+1 is preferred, clearly suggesting
that some information about the sub-structure in the pixel can be extracted from the observations (apart
from the magnetic flux density that can be obtained from the WEAKF approximation).
which might be surprising for such a poor spectral sampling.
Note that the weak-field approximation allows one to extract very simple quantities from the observables, 
without allowing for an \emph{explicit} substructure inside the pixel. We point out that it might be possible
to find more complex models but with a reduced number of parameters (smaller than ME1+1) so that
it can be preferred by data. The only way to decide on this is to calculate the value of the 
evidence and compute the evidence ratio.

\subsection{Simpler proxies}
Given that calculating a reliable estimation of the evidence is computationally very demanding, 
it is of interest to compare it with simpler proxies used for model comparison. The property
of such proxies is that they can be calculated very fast and it is not necessary to 
perform the multidimensional integral of the evidence. Two of the simple, routinely used proxies
are the Bayesian Information Criterion \citep[BIC;][]{schwarz_bic78} and the Akaike Information
Criterion \citep[AIC;][]{akaike_aic78}. Both methods, which are based on the crude approximation 
of gaussianity of the posterior with respect to the model parameters, are extremely simple to calculate:
\begin{eqnarray}
\mathrm{BIC} &=& -2 \ln \mathcal{L}_\mathrm{max} + k \ln N \\
\mathrm{AIC} &=& -2 \ln \mathcal{L}_\mathrm{max} + 2k,
\end{eqnarray}
where $k$ is the number of free parameters of the model, $\mathcal{L}_\mathrm{max}$ is the peak value
of the likelihood (at the least-squares solution if flat priors are used) and $N$ is the
number of observed points. In the case of a Gaussian likelihood, they transform to:
\begin{eqnarray}
\mathrm{BIC} &=& \chi^2_\mathrm{min} + k \ln N \nonumber \\
\mathrm{AIC} &=& \chi^2_\mathrm{min} + 2k,
\label{eq:bic}
\end{eqnarray}
which can be readly calculated for standard inversion methods based on a least-squares
minimization, using the set of parameters $\hat{\thetabold}$ that minimize
the $\chi^2$:
\begin{equation}
\chi^2_\mathrm{min} = \sum_{j=1}^M \left( \frac{y_j(\hat{\thetabold})-d_j}{\sigma_j} \right)^2.
\end{equation}
One of the fundamental problems of these criteria (apart from the 
assumption of gaussianity of the posterior) is that they penalize
all parameters equally, not taking into account situations in which data does not
constrain some parameters.

The computed values of the BIC are shown in Table \ref{tab:bic}. The model with the
smallest value of the BIC is the preferred one, contrary to what happens with the
evidence. This model is indicated in bold red when it gives the same result using
the Bayesian evidence and in blue when it gives a different result.
When comparing two models, we have verified that more than 80\% of the time
the BIC picks up the same model selected by the evidence ratio when dealing with
Hinode observations, while this value increases to $\sim$90\% when focusing on IMaX
profiles. The success rate for selecting the best model using the BIC (as compared with the
fully Bayesian case) goes down to 73\%. We consider this an indication that the
BIC is a very good proxy for the Bayesian evidence.
The AIC performs similarly, with BIC being slightly more robust.
For instance, the LINGR2 model is preferred by all model comparison methods for the
profile associated to Class 4. On the contrary, the weak-field approximation is preferred
for Class 9 according to the evidence ratio, while the more complex NOGR2 model is the one of choice
according to BIC. Note that, according to the evidence ratio, this is the next preferred model
in the hierarchy. 

In any case, given the relatively large success rate of BIC for comparison of
two models, we suggest anyone carrying out standard inversions to compute the
value of the BIC for the selected model. This facilitates model comparison in the
future and is able to select the more probable model in a comparison of two with 
$\sim$80\% confidence if our results are used as a calibration. Another application
of interest of the proxy is to estimate the minimum number of wavelength points used to
sampled the Stokes profiles when observing an unresolved magnetic structure. If
one confronts a ME model with one magnetic component and a ME1+1 model to obtain information 
about the filling-factor, the ME1+1 model will be preferred when BIC(ME1+1) is larger 
than BIC(ME). With the estimated values of the $\chi^2$, one can infer the number of wavelength
points to prefer ME1+1.

% Another quantity of interest for simple model comparison that avoids these problems
% is the Deviance Information Criterion \citep{spiegelhalter_complexity02}. 
% This quantity gives an estimation of the number
% of model parameters that can be effectively constrained with the current data and can be
% used to discriminate between two models with roughly the same evidence. Under some
% approximations explained in detail in \cite{spiegelhalter_complexity02} and \cite{trotta08}, and
% assuming that the errors corrupting the observations follow a Gaussian distribution, the
% Bayesian complexity is expressed as:
% \begin{equation}
% \mathcal{C}_b = \overline{\chi^2(\thetabold)} - \chi^2(\hat{\thetabold}).
% \end{equation}
% The first term is the mean $\chi^2$ taken over the posterior distribution, i.e.:
% \begin{eqnarray}
% \overline{\chi^2(\thetabold)} &=& \int \mathrm{d}\thetabold_i p(\thetabold_i|D,\mathcal{M}_i) \ln p(D|\thetabold_i,\mathcal{M}_i) \nonumber \\
% &=& -\frac{1}{2} \int \mathrm{d}\thetabold_i p(\thetabold_i|D,\mathcal{M}_i) \chi^2(\thetabold_i),
% \end{eqnarray}
% where the last step comes from the presence of Gaussian noise, since the likelihood fulfills 
% $\chi^2 = -2 \ln p(D|\thetabold_i,\mathcal{M}_i)$. The second term is just the $\chi^2$ evaluated
% at the estimated value of the parameteres, that we choose to be the maximum a-posteriori values.

%%%%%%%%%%%%%%%%%%%%%%%%%%%%%%%%%%%%%%%%%%%%%%%%%%%%%%%%%
%%%%%%%%%%%%%%%%%%%%%%%%%%%%%%%%%%%%%%%%%%%%%%%%%%%%%%%%%

\section{Conclusions}
We have presented the first quantitative Bayesian comparison of models used for the
interpretation of observed Stokes profiles. Our results suggest that there is not
a single model that is suitable for explaining different Stokes profiles
in the quiet Sun and in active regions. In essence, the selected model in each case
depends on the amount of information encoded in the observations. Simpler models
are preferred when the SNR is low or when the spectral sampling is poor because this
information is diluted by the noise. Even if the underlying physics is very complex
and is producing inherently very complicated Stokes profiles, the presence of noise destroys the
information and a simple model is enough to explain them.
More complex models are favored when the SNR is large (especially if this is the case
for the four Stokes parameters simultaneously) because minute variations of the
shape of the Stokes parameters have to be fitted. We stress again that the higher 
SNR is the responsible for the need of increased complexity just
because it does not suppress the important details of the Stokes profiles that
arise in complex atmospheres (asymmetries, several components, etc.). Then, we conclude
that the complexity of the observed Stokes profiles is the main driver for favoring 
more elaborate models.

Given that the Bayesian evidence is computationally heavy, we
suggest to use the BIC as a trivial output of any inversion code. This will facilitate
a more quantitative model comparison in the future, if used as a proxy of the Bayesian evidence.
Additionally, the good behavior of the BIC can be used to develop an metainversion scheme 
in which any of the Eqs. (\ref{eq:bic}) is minimized modifying the values of the free parameters of the models
and the models themselves. The result would be a good approximation to the best model that
is allowed by the data.

\acknowledgements
We thank C. Beck and B. Lites for useful suggestions. Financial support by the 
Spanish Ministry of Science and Innovation through projects AYA2010-18029 (Solar Magnetism and Astrophysical 
Spectropolarimetry) and Consolider-Ingenio 2010 CSD2009-00038 is gratefully acknowledged.
This research has greatly beneﬁted from discussions that were held at the 
International Space Science Institute (ISSI) in Bern (Switzerland) in February 2010 
as part of the International Working group \textit{Extracting information from spectropolarimetric 
observations: comparison of inversion codes}.

\appendix

\section{Appendix}
\label{sec:appendix}
For the sake of clarity, we write the explicit expressions we have used in each model for the
computation of the Stokes profiles.

\subsection{Weak-field approximation}
In the weak-field approximation, these are the relations between Stokes $Q$, $U$ and $V$ and
Stokes $I$ \citep{landi_landolfi04}:
\begin{eqnarray}
V(\lambda) &=& -4.67 \times 10^{-13} \bar{g} \lambda^2 \Bpar \didlnoi \nonumber \\
Q(\lambda) &=& -5.45 \times 10^{-26} \bar{G} \lambda^4 \Bperp^2 \coschi \didltwonoi \nonumber \\
U(\lambda) &=& -5.45 \times 10^{-26} \bar{G} \lambda^4 \Bperp^2 \sinchi \didltwonoi,
\label{eq:circ_lin_pol}
\end{eqnarray}
with the wavelength $\lambda$ in \AA\ and the components of the magnetic field vector measured in G.
The factor $\bar{g}$ is the effective Land\'e factor and $\bar{G}$ is the
equivalent for linear polarization \citep[e.g.,][]{landi_landolfi04}. The relations at
second order for Stokes $Q$ and $U$ are only valid for non-saturated lines. In our modified weak-field
approximation, Stokes $I$ is given by Eq. (\ref{eq:stokesi_wf}) and its derivatives can be
computed analytically.

\subsection{Milne-Eddington solution}
In the Milne-Eddington approximation, the emergent Stokes profiles normalized to the continuum
intensity are given by Eqs. (9.110) of \cite{landi_landolfi04}, that we rewrite here:
\begin{eqnarray}
\frac{I(\mu)}{I_c} &=& \frac{1}{1+\beta \mu} \left\{ 1+\beta \mu \Delta^{-1} (1+k_I) \left[ (1+k_I)^2 + f_Q^2 + f_U^2 + f_V^2 \right] \right\} \nonumber \\
\frac{Q(\mu)}{I_c} &=& -\frac{\beta \mu}{1+\beta \mu} \Delta^{-1} \left\{(1+k_I)^2 k_Q - (1+k_I)(k_U f_V - k_V f_U) + f_Q(k_Q f_Q + k_U f_U + k_V f_V) \right\} \nonumber \\
\frac{U(\mu)}{I_c} &=& -\frac{\beta \mu}{1+\beta \mu} \Delta^{-1} \left\{(1+k_I)^2 k_U - (1+k_I)(k_V f_Q - k_Q f_V) + f_U(k_Q f_Q + k_U f_U + k_V f_V) \right\} \nonumber \\
\frac{V(\mu)}{I_c} &=& -\frac{\beta \mu}{1+\beta \mu} \Delta^{-1} \left\{(1+k_I)^2 k_V - (1+k_I)(k_Q f_U - k_U f_Q) + f_V(k_Q f_Q + k_U f_U + k_V f_V) \right\},
\end{eqnarray}
where
\begin{equation}
\Delta = (1+k_I)^4 + (1+k_I)^2 (f_Q^2+f_U^2+f_V^2-k_Q^2-k_U^2-k_V^2) - (k_Q f_Q + k_U f_U + k_V f_V)^2.
\end{equation}
The $k_i$ with $i=\{I,Q,U,V\}$ and $f_i$ with $i=\{Q,U,V\}$ are the elements of the propagation matrix,
as defined in Eq. (9.39) of \cite{landi_landolfi04}.

\section{Maximum a-posteriori values}
\label{sec:app_map}
Table \ref{tab:map_table} shows, for all models except the weak-field approximation, the maximum a-posteriori
values of: the field strength ($B$), field inclination ($\theta_B$), filling factor ($f$) and the magnetic flux density ($\phi$). 
In these models, the magnetic flux density is computed as $\phi=f B_1 \cos({[\theta_B]}_1)+(1-f) B_2 \cos({[\theta_B]}_2)$.
Given the special nature of the weak-field model, we only tabulated the magnetic flux density, which is given
by $B_\parallel$ since there is only one component. Since we use flat priors, the tabulated values
coincide with those one would obtain using a standard least-squares fitting of the observed profiles. 
For the models with two magnetic components, we show the value of $B$ and $\theta_B$ in both
components. For the models with gradients along the line-of-sight, we tabulate the
value at $\log \tau_{5000}=-2$, as a representation of the conditions in the line formation region. We have
not shown the error bars to avoid crowding.
\begin{table*}
\caption{Maximum a-posteriori values for some parameters.}
\centering
\scriptsize
\begin{tabular}{ccccccccc}
$\phi$ [Mx cm$^{-2}$] & WEAKF & ME1+1 & ME2 & NOGR1+1 & NOGR2 & LINGR1+1 & LINGR2 \\
\hline\hline
Class 0 &  $  -72.54$ &  $  -62.39$ &  $  -67.83$ &  $ -122.52$ &  $ -145.91$ &  $ -197.62$ &  $ -153.47$ \\ 
Class 1 &  $   26.71$ &  $   30.89$ &  $   26.67$ &  $   40.37$ &  $  -38.94$ &  $   65.46$ &  $  125.00$ \\ 
Class 2 &  $   17.80$ &  $   14.02$ &  $   16.97$ &  $   22.64$ &  $   18.65$ &  $   44.93$ &  $    3.74$ \\ 
Class 4 &  $  676.25$ &  $  734.28$ &  $  708.20$ &  $  501.82$ &  $  615.53$ &  $  514.85$ &  $  601.46$ \\ 
Class 9 &  $   14.50$ &  $    3.90$ &  $   19.52$ &  $   24.51$ &  $   19.62$ &  $   55.32$ &  $   20.64$ \\ 
Class 11 &  $   34.95$ &  $   25.71$ &  $   36.91$ &  $   44.36$ &  $   47.33$ &  $   97.59$ &  $   32.41$ \\ 
Class 17 &  $   16.54$ &  $   15.62$ &  $   19.12$ &  $   23.89$ &  $   27.67$ &  $   46.00$ &  $   23.82$ \\ 
Class 25 &  $   -9.41$ &  $    0.11$ &  $  -15.85$ &  $  -10.27$ &  $   -1.97$ &  $  -25.50$ &  $   -2.25$ \\ 
Class 34 &  $    6.28$ &  $    8.28$ &  $    9.36$ &  $   14.51$ &  $   10.73$ &  $   31.93$ &  $  -12.75$ \\ 
Penumbra &  $  697.89$ &  $  725.43$ &  $  719.37$ &  $  543.02$ &  $  782.60$ &  $  839.17$ &  $  819.95$ \\ 
Umbra &  $  999.86$ &  $ 1298.82$ &  $ 2532.40$ &  $ 2279.34$ &  $ 2899.99$ &  $ 1703.17$ &  $ 2684.40$ \\ 
IMaX1 &  $   -6.11$ &  $  -11.44$ &  $   -0.18$ &  $    4.64$ &  $   68.47$ &  $ -251.23$ &  $-1034.43$ \\ 
IMaX2 &  $   43.97$ &  $   51.38$ &  $  -31.95$ &  $  135.27$ &  $    7.88$ &  $  465.60$ &  $   15.86$ \\ 
IMaX3 &  $  -18.86$ &  $  -14.42$ &  $  -77.15$ &  $ -281.55$ &  $ -748.89$ &  $ -376.38$ &  $ -514.77$ \\ 
IMaX4 &  $   -1.39$ &  $    1.42$ &  $   -1.99$ &  $  -34.24$ &  $  125.99$ &  $    0.21$ &  $ 1020.42$ \\ 
\hline
$B$ [G] &  &  &  &  &  &  &  \\
\hline\hline
Class 0 &  $-$ &  $ 1285.24$ &  $104.79/234.61$ &  $  228.00$ &  $247.00/82.20$ &  $  367.00$ &  $357.00/47.90$ \\ 
Class 1 &  $-$ &  $  114.62$ &  $59.03/81.74$ &  $   81.10$ &  $52.00/96.70$ &  $  143.00$ &  $68.10/229.00$ \\ 
Class 2 &  $-$ &  $   49.73$ &  $58.35/54.31$ &  $   58.20$ &  $52.20/48.70$ &  $  103.00$ &  $36.40/178.00$ \\ 
Class 4 &  $-$ &  $  932.46$ &  $617.19/864.72$ &  $  977.00$ &  $538.00/819.00$ &  $  814.00$ &  $844.00/567.00$ \\ 
Class 9 &  $-$ &  $   29.80$ &  $41.18/42.59$ &  $   58.60$ &  $79.00/44.30$ &  $  102.00$ &  $106.00/29.90$ \\ 
Class 11 &  $-$ &  $  218.74$ &  $78.82/179.61$ &  $  207.00$ &  $167.00/161.00$ &  $  395.00$ &  $155.00/69.30$ \\ 
Class 17 &  $-$ &  $  223.00$ &  $93.53/52.60$ &  $  157.00$ &  $41.60/151.00$ &  $  248.00$ &  $134.00/57.30$ \\ 
Class 25 &  $-$ &  $   15.98$ &  $58.29/48.57$ &  $   95.90$ &  $82.00/54.20$ &  $  191.00$ &  $213.00/57.00$ \\ 
Class 34 &  $-$ &  $  218.70$ &  $206.35/198.37$ &  $  298.00$ &  $126.00/337.00$ &  $  488.00$ &  $545.00/136.00$ \\ 
Penumbra &  $-$ &  $ 1337.04$ &  $1299.85/1337.64$ &  $ 1320.00$ &  $1480.00/930.00$ &  $ 1430.00$ &  $1400.00/1270.00$ \\ 
Umbra &  $-$ &  $ 2927.44$ &  $2888.75/2253.19$ &  $ 3030.00$ &  $2630.00/3810.00$ &  $ 2480.00$ &  $3800.00/2110.00$ \\ 
IMaX1 &  $-$ &  $ 1369.61$ &  $59.87/97.49$ &  $ 2650.00$ &  $1870.00/62.70$ &  $ 2230.00$ &  $2170.00/82.90$ \\ 
IMaX2 &  $-$ &  $  963.44$ &  $166.10/34.36$ &  $  381.00$ &  $259.00/41.20$ &  $  798.00$ &  $43.00/177.00$ \\ 
IMaX3 &  $-$ &  $  435.38$ &  $140.52/99.94$ &  $ 1530.00$ &  $84.60/3420.00$ &  $ 1530.00$ &  $2260.00/88.70$ \\ 
IMaX4 &  $-$ &  $  398.50$ &  $1.93/51.77$ &  $  572.00$ &  $13.30/3090.00$ &  $ 1890.00$ &  $102.00/3260.00$ \\ 
\hline
$\theta_B$ [deg] &  &  &  &  &  &  & \\
\hline\hline
Class 0 &  $-$ &  $  178.52$ &  $117.86/137.50$ &  $  149.30$ &  $158.60/131.60$ &  $  156.40$ &  $139.50/130.70$ \\ 
Class 1 &  $-$ &  $   64.05$ &  $103.24/63.14$ &  $   40.70$ &  $61.50/169.80$ &  $   40.30$ &  $54.10/22.80$ \\ 
Class 2 &  $-$ &  $   54.85$ &  $69.66/87.38$ &  $   52.00$ &  $65.90/72.50$ &  $   40.40$ &  $67.80/92.90$ \\ 
Class 4 &  $-$ &  $   24.34$ &  $25.29/25.75$ &  $   15.00$ &  $43.40/17.80$ &  $   23.80$ &  $23.60/23.10$ \\ 
Class 9 &  $-$ &  $    3.57$ &  $64.72/46.25$ &  $   48.30$ &  $71.20/74.90$ &  $   20.00$ &  $75.20/63.60$ \\ 
Class 11 &  $-$ &  $   75.29$ &  $51.24/79.07$ &  $   71.60$ &  $80.20/69.10$ &  $   66.20$ &  $80.00/58.10$ \\ 
Class 17 &  $-$ &  $   78.38$ &  $79.37/59.38$ &  $   72.90$ &  $76.20/75.10$ &  $   70.10$ &  $75.50/77.50$ \\ 
Class 25 &  $-$ &  $   88.13$ &  $98.26/125.90$ &  $  104.90$ &  $84.90/102.10$ &  $  104.30$ &  $86.00/104.20$ \\ 
Class 34 &  $-$ &  $   77.46$ &  $77.68/89.36$ &  $   84.10$ &  $87.60/87.10$ &  $   82.20$ &  $93.90/87.10$ \\ 
Penumbra &  $-$ &  $   51.47$ &  $53.44/64.39$ &  $   53.00$ &  $49.80/59.30$ &  $   50.10$ &  $50.50/59.60$ \\ 
Umbra &  $-$ &  $   17.80$ &  $21.52/21.43$ &  $   15.90$ &  $26.20/27.20$ &  $   16.80$ &  $27.70/26.50$ \\ 
IMaX1 &  $-$ &  $  100.67$ &  $93.36/28.90$ &  $   89.10$ &  $79.60/96.90$ &  $  109.70$ &  $186.10/94.70$ \\ 
IMaX2 &  $-$ &  $   13.36$ &  $137.65/128.97$ &  $   47.20$ &  $112.50/36.90$ &  $   19.80$ &  $31.50/111.70$ \\ 
IMaX3 &  $-$ &  $  103.04$ &  $137.34/138.23$ &  $  111.70$ &  $97.00/152.30$ &  $  121.40$ &  $143.80/96.40$ \\ 
IMaX4 &  $-$ &  $   75.73$ &  $140.44/126.82$ &  $   96.70$ &  $89.10/40.50$ &  $   89.20$ &  $92.10/34.00$ \\ 
\hline
$f$ &  &  &  &  &  &  &  \\
\hline\hline
Class 0 &  $-$ &  $    0.05$ &  $    0.85$ &  $    0.62$ &  $    0.52$ &  $    0.59$ &  $    0.51$ \\ 
Class 1 &  $-$ &  $    0.62$ &  $    0.20$ &  $    0.66$ &  $    0.47$ &  $    0.60$ &  $    0.50$ \\ 
Class 2 &  $-$ &  $    0.49$ &  $    0.81$ &  $    0.63$ &  $    0.60$ &  $    0.57$ &  $    0.56$ \\ 
Class 4 &  $-$ &  $    0.86$ &  $    0.32$ &  $    0.53$ &  $    0.42$ &  $    0.69$ &  $    0.32$ \\ 
Class 9 &  $-$ &  $    0.13$ &  $    0.84$ &  $    0.63$ &  $    0.58$ &  $    0.58$ &  $    0.53$ \\ 
Class 11 &  $-$ &  $    0.46$ &  $    0.19$ &  $    0.68$ &  $    0.35$ &  $    0.61$ &  $    0.43$ \\ 
Class 17 &  $-$ &  $    0.35$ &  $    0.80$ &  $    0.52$ &  $    0.39$ &  $    0.54$ &  $    0.54$ \\ 
Class 25 &  $-$ &  $    0.21$ &  $    0.63$ &  $    0.42$ &  $    0.50$ &  $    0.54$ &  $    0.41$ \\ 
Class 34 &  $-$ &  $    0.17$ &  $    0.17$ &  $    0.47$ &  $    0.54$ &  $    0.48$ &  $    0.45$ \\ 
Penumbra &  $-$ &  $    0.87$ &  $    0.72$ &  $    0.68$ &  $    0.64$ &  $    0.91$ &  $    0.72$ \\ 
Umbra &  $-$ &  $    0.47$ &  $    0.74$ &  $    0.78$ &  $    0.47$ &  $    0.72$ &  $    0.54$ \\ 
IMaX1 &  $-$ &  $    0.05$ &  $    0.96$ &  $    0.11$ &  $    0.22$ &  $    0.33$ &  $    0.48$ \\ 
IMaX2 &  $-$ &  $    0.05$ &  $    0.10$ &  $    0.52$ &  $    0.19$ &  $    0.62$ &  $    0.80$ \\ 
IMaX3 &  $-$ &  $    0.15$ &  $    0.09$ &  $    0.50$ &  $    0.76$ &  $    0.47$ &  $    0.28$ \\ 
IMaX4 &  $-$ &  $    0.01$ &  $    0.98$ &  $    0.51$ &  $    0.95$ &  $    0.01$ &  $    0.62$
\end{tabular}
\label{tab:map_table}
\end{table*}

% \bibliographystyle{apj}
% \bibliography{/scratch/Dropbox/biblio}

\end{document}